\documentclass[fleqn]{llncs} 
\usepackage{datbb} 

\title{Dormancy-Aware Timed Branching Bisimilarity}
\subtitle{With an Application to Communication Protocol Analysis}
\author{C.A. Middelburg \\
       {\small ORCID: \url{https://orcid.org/0000-0002-8725-0197}}}
\institute{Informatics Institute, Faculty of Science, 
           University of Amsterdam \\
           Science Park~904, 1098~XH Amsterdam, the Netherlands \\
           \email{C.A.Middelburg@uva.nl}}
  
\begin{document}
\maketitle

\begin{abstract}
A variant of the standard notion of branching bisimilarity for processes 
with discrete relative timing is proposed which is coarser than the 
standard notion.
Using a version of ACP (Algebra of Communicating Processes) with 
abstraction for processes with discrete relative timing, it is shown 
that the proposed variant allows of both the functional correctness and 
the performance properties of the PAR (Positive Acknowledgement with 
Retransmission) protocol to be analyzed.
In the version of \ACP\ concerned, the difference between the standard 
notion of branching bisimilarity and its proposed variant is 
characterized by a single axiom schema.
\\[1.5ex]
\\[1.5ex]
{\sl 1998 ACM Computing Classification:} 
C.2.2, D.1.3, D.2.4, F.1.2, F.3.1.
\end{abstract}

\section{Introduction}
\label{sect-intro}

The axiom systems of versions of \ACP\ (Algebra of Communicating 
Processes) with abstraction for processes with discrete relative timing 
are based on the notion of branching bisimilarity from~\cite{GW96a} 
adapted to processes with discrete relative timing (see 
e.g.~\cite{BB95b,BBR98a,BM02a}).
The experience with analyzing the PAR (Positive Acknowledgement with 
Retransmission) protocol~\cite[Section~3.3]{Tan81a} during the writing 
of~\cite{BM02a} triggered a quest for a variant of that notion of 
branching bisimilarity that is coarser.
In this paper, such a variant is proposed.

The PAR protocol is a communication protocol which is based on time-outs
of positive acknowledgements.
Timing is essential for the correct behaviour of the PAR protocol: 
the protocol behaves only correctly if the time-out time of the sender, 
i.e.\ the time after which it retransmits a datum for which it is still
awaiting an acknowledgement, is longer than the time that a complete
protocol cycle takes. 
There have been attempts to describe this protocol using process 
algebra without timing.
In those attempts, see e.g.~\cite{Vaa90a}, it is excluded that the 
sender times out too early by inhibiting a time-out so long as it does 
not lead to deadlock.
This boils down to assuming a sender that is somehow able to determine,
when it is waiting for an acknowledgement, whether the receiver and the
channels used for communication are at rest --- this is nicely 
illustrated in~\cite{Bru95a}.
Such attempts resort to assumptions of which it is unlikely that they 
can be fulfilled by actual communication protocols. 
% I will not go into the attempts where the premature time-out of the 
% sender is not excluded at all.

In~\cite{BM02a}, a version of \ACP\ with abstraction for processes with 
discrete relative timing is used to describe the PAR protocol and to 
show that the protocol behaves correctly if the time-out time of the 
sender is longer than the time that a complete protocol cycle takes.
In order to achieve this result, it is necessary to take the timing of 
actions into account in all calculations that must be done before 
abstraction from the actions that should be regarded as internal can 
be applied.
From the point where abstraction from those actions can be applied, the
timing of actions is no longer relevant.
Actually, many internal actions can only be removed after abstraction
from the timing of actions.

It would facilitate analysis of performance properties if most internal 
actions could be removed without preceding abstraction from the timing 
of actions.
The axioms used for the removal of internal actions are based on the 
notion of branching bisimilarity for processes with discrete relative 
timing introduced in~\cite{BB95b,BBR98a,BM02a}.
What is needed in the case of performance analysis, is a coarser 
equivalence relation.
This is the main motivation for the proposal in this paper of a coarser 
equivalence relation, with a plausible rationale, which still coincides 
with the original version of branching bisimilarity from~\cite{GW96a} in 
the case without timing.

In~\cite{BMR02b}, a first attempt has been made to define this coarser
version.
The attempt was unsatisfactory.
The definition in that paper was in fact complicated to such an extent 
that it led to errors in the paper that were not spotted at the time.
Like the definition of the standard version, the definition concerned 
refers to a two-phase operational semantics of a version of \ACP\ with 
abstraction for processes with discrete relative timing.
In this paper, a definition will be given that refers to a time-stamped 
operational semantics instead.
It turns out that taking a time-stamped operational semantics as the 
basis results in a definition that is far less complicated than the one 
from~\cite{BMR02b}.

To give a first indication of the usefulness of the proposed equivalence,
it is also shown in this paper that in the case of the PAR protocol all 
internal actions that hinder performance analysis can be removed without 
preceding abstraction from the timing of actions if we use the axioms 
based on the proposed equivalence.
Because of the plausible rationale behind this equivalence that is given 
in this paper, it is likely that the usefulness of the proposed 
equivalence is not restricted to the PAR protocol.

The contributions of this paper are not only the definition of a variant 
of the standard notion of branching bisimilarity for processes with 
discrete relative timing and the devising of an axiom schema 
characterizing the difference between the standard notion and the 
variant, but also 
(a)~the definition of both a two-phase structural operational semantics 
and a time-stamped structural operational semantics of a version of 
\ACP\ with abstraction for processes with discrete relative timing, 
(b)~two equivalent definitions of the standard notion of branching 
bisimilarity for processes with discrete relative timing, one referring 
to the two-phase semantics and one referring to the time-stamped 
semantics, and 
(c)~the extension of the version of \ACP\ being considered with 
unrestricted nested guarded recursion.
The version of \ACP\ being considered is arguably the core of the 
versions presented earlier in~\cite{BB95b,BBR98a,BM02a}.
To the best of my knowledge, time-stamped structural operational 
semantics of process algebras for processes with relative timing and 
process algebras with unrestricted nested guarded recursion have not 
been considered in the computer science literature before.

The structure of this paper is as follows. 
First, the version of \ACP\ with abstraction for processes with 
discrete relative timing used in this paper is presented 
(Section~\ref{sect-prelims}).
Next, this version is used to formally specify the PAR protocol and to 
analyze it (Section~\ref{sect-par-protocol}).
Then, the standard notion of branching bisimilarity for processes with 
discrete relative timing is defined in two ways
(Section~\ref{sect-standard-bisim}).
Following this, the variant of the standard notion referred to above is 
introduced and an axiom schema that characterizes the difference between 
the two notions is given (Section~\ref{sect-coarser-eqv}).  
After that, the analysis is revisited using that axiom schema
(Section~\ref{sect-analysis-3}).
Finally, some concluding remarks are made 
(Section~\ref{sect-conclusion}).

In~\cite{BM02a}, a coherent collection of four process algebras with 
timing, each dealing with timing in a different way, is presented. 
The time scale on which the time is measured is either discrete or 
continuous, and the timing of actions is either relative or absolute.
In the current paper, a minor variant of the process algebra with 
discrete relative timing from that collection is used.
Various constants and operators of that process algebra have 
counterparts in the other process algebras from the collection.
In~\cite{BM02a}, a notational distinction is made between a constant or 
operator of one process algebra and its counterparts in other process 
algebras, by means of different decorations of a common symbol, if they 
should not be identified in case process algebras are compared.
In this paper, the decorations are omitted because they can safely be
omitted so long as a single process algebra is used.

\section{Process Algebraic Preliminaries}
\label{sect-prelims}

The version of \ACP\ used in this paper can be considered the core of 
\ACP$^\drt_\tau$, the process algebra with abstraction for processes 
with discrete relative timing from~\cite{BM02a}, extended with nested 
guarded recursion.
To distinguish it from \ACP$^\drt_\tau$, the core of \ACP$^\drt_\tau$ is 
denoted by \ACP$_\drt^\tau$ in this paper.
In this section, \ACP$_\drt^\tau$ and its extension with nested guarded 
recursion are introduced.
It focuses on the signatures and axiom systems of these algebraic 
theories.
The congruence with respect to which closed substitution instances of 
the equations and conditional equations from the axiom systems are valid
is introduced in Section~\ref{sect-standard-bisim}.

\subsection{\ACP$_\drt^\tau$}
\label{subsect-acpdrt}

In the case of \ACP$_\drt^\tau$, timing is relative to the time at which 
the preceding action is performed and time is measured on a discrete 
time scale.
Measuring time on a discrete time scale means that time is divided into
time slices and timing of actions is done with respect to the time
slices in which they are performed.
Roughly speaking, \ACP$_\drt^\tau$ is \ACP$^\tau$ extended with two 
operators to deal with timing: one-time-slice delay and 
current-time-slice time-out.
The former operator is a basic one and the latter operator is a useful 
auxiliary one.

In \ACP$_\drt^\tau$, it is assumed that a fixed but arbitrary finite set 
$\Act$ of \emph{basic actions}, with $\tau,\dead \not\in \Act$, and 
a fixed but arbitrary commutative and associative \emph{communication} 
function 
$\funct{\commf}
 {(\Act \union \set{\tau,\dead}) \x (\Act \union \set{\tau,\dead})}
 {(\Act \union \set{\tau,\dead})}$, 
such that $\commf(\tau,a) = \dead$ and $\commf(\dead,a) = \dead$
for all $a \in \Act \union \set{\tau,\dead}$, have been given.
Basic actions are taken as atomic processes.
The function $\commf$ is regarded to give the result of synchronously
performing any two basic actions for which this is possible, and to be 
$\dead$ otherwise.
Henceforth, we write $\Actt$ for $\Act \union \set{\tau}$ and $\Acttd$ 
for $\Act \union \set{\tau,\dead}$.

The signature of the algebraic theory \ACP$_\drt^\tau$ consists of the 
following constants and operators: 
\begin{itemize}
\item
for each $a \in \Act$, 
the \emph{undelayable action} constant $\cts{a}$\,;
\item
the \emph{undelayable silent step} constant $\cts{\tau}$;
\item
the \emph{undelayable deadlock} constant $\cts{\dead}$\,;
\item
the binary \emph{alternative composition} operator $\altc$\,;
\item
the binary \emph{sequential composition} operator $\seqc$\,;
\item
the unary \emph{one-time-slice delay} operator $\delay$\,;
\item
the binary \emph{parallel composition} operator $\parc$\,;
\item
the binary \emph{left merge} operator $\leftm$\,;
\item
the binary \emph{communication merge} operator $\commm$\,;
\item
for each $H \subseteq \Act$, 
the unary \emph{encapsulation} operator $\encap{H}$\,;
\item
for each $I \subseteq \Act$, 
the unary \emph{abstraction} operator $\abstr{I}$\,;
\item
the unary \emph{current-time-slice time-out} operator $\timeout$\,.
\end{itemize}
It is assumed that there is a countably infinite set $\cX$ of variables, 
which contains $x$, $y$ and $z$.
Terms over the signature of \ACP$_\drt^\tau$ are built as usual.
Infix notation is used for the binary operators.
The need to use parentheses is reduced by using the associativity of 
the operators $\altc$, $\seqc$ and $\parc$ and the following precedence 
conventions: the operator $\seqc$ binds stronger than all other binary 
operators and the operator $\altc$ binds weaker than all other binary 
operators.
Henceforth, terms over the signature of \ACP$_\drt^\tau$ are also called 
\ACP$_\drt^\tau$ terms.
This terminology will also  be used for terms over the signature of 
restrictions and extensions of \ACP$_\drt^\tau$.

The constants of \ACP$_\drt^\tau$ can be explained as follows 
($a \in \Act$):
\begin{itemize}
\item
$\cts{a}$ denotes the process that performs the observable action $a$ in 
the current time slice and after that immediately terminates 
successfully;
\item
$\cts{\tau}$ denotes the process that performs an unobservable action in 
the current time slice and after that immediately terminates 
successfully;
\item
$\cts{\dead}$ denotes the process that is neither capable of performing 
any action in the current time slice nor capable of idling till the next 
time slice.
\end{itemize}
Let $t$ and $t'$ be closed \ACP$_\drt^\tau$ terms denoting processes $p$ 
and $p'$, and let $H,I \subseteq \Act$.
Then the operators of \ACP$_\drt^\tau$ can be explained as follows:
\begin{itemize}
\item
$t \altc t'$ denotes the process that behaves either as $p$ or as $p'$, 
where the choice between the two is resolved at the instant that one of 
them performs its first action, and not before;
\item
$t \seqc t'$ denotes the process that first behaves as $p$ and following 
successful termination of $p$ behaves as $p'$;
\item
$\delay(t)$ denotes the process that idles till the next time slice 
and then behaves as $p$;
\item
$t \parc t'$ denotes the process that behaves as $p$ and $p'$ in 
parallel, by which is meant that, each time an action is performed, 
either a next action of $p$ is performed or a next action of $p'$ is 
performed or a next action of $p$ and a next action of $p'$ are 
performed synchronously;
\item
$t \leftm t'$ denotes the same process as $t \parc t'$, except that it 
starts with performing an action of $p$;
\item
$t \commm t'$ denotes the same process as $t \parc t'$, except that it 
starts with performing an action of $p$ and an action of $p'$ 
synchronously;
\item
$\encap{H}(t)$ denotes the process that behaves the same as $p$, except 
that it keeps $p$ from performing actions in $H$;
\item
$\abstr{I}(t)$ denotes the process that behaves the same as $p$, except 
that actions in $I$ are turned into unobservable actions;
\item
$\timeout(t)$, denotes the process that behaves the same as $p$, 
except that it keeps $p$ from idling till the next time slice.
\end{itemize}

The axiom system of \ACP$_\drt^\tau$ consists of the equations given in
Table~\ref{axioms-ACPdrtt}.
\begin{table}[!t]
\caption{Axiom system of \ACP$_\drt^\tau$}
\label{axioms-ACPdrtt}
\begin{eqntbl}
\begin{array}[t]{@{}c@{}}
\begin{axcol}
x \altc y = y \altc x                                  & \axiom{A1}   \\
(x \altc y) \altc z = x \altc (y \altc z)              & \axiom{A2}   \\
x \altc x = x                                          & \axiom{A3}   \\
(x \altc y) \seqc z = x \seqc z \altc y\seqc z         & \axiom{A4}   \\
(x \seqc y) \seqc z = x \seqc (y \seqc z)              & \axiom{A5}   \\
x \altc \cts{\dead} = x                                & \axiom{A6DR} \\
\cts{\dead} \seqc x = \cts{\dead}                      & \axiom{A7DR}
\\ {} \\
\delay(x) \altc \delay(y) = \delay(x \altc y)          & \axiom{DRT1} \\
\delay(x) \seqc y = \delay(x \seqc y)                  & \axiom{DRT2} 
\\ {} \\
\encap{H}(\cts{a}) = \cts{a}\;\; \mif a \not\in H      & \axiom{D1DR} \\
\encap{H}(\cts{a}) = \cts{\dead}\;\; \mif a \in H      & \axiom{D2DR} \\
\encap{H}(x \altc y) = \encap{H}(x) \altc \encap{H}(y) & \axiom{D3}   \\
\encap{H}(x \seqc y) = \encap{H}(x) \seqc \encap{H}(y) & \axiom{D4}   \\
\encap{H}(\delay(x)) = \delay(\encap{H}(x))            & \axiom{DRD}  
\\ {} \\
\abstr{I}(\cts{a}) = \cts{a}\;\; \mif a \not\in I      & \axiom{TI1DR}\\
\abstr{I}(\cts{a}) = \cts{\tau}\;\; \mif a \in I       & \axiom{TI2DR}\\
\abstr{I}(x \altc y) = \abstr{I}(x) \altc \abstr{I}(y) & \axiom{TI3}  \\
\abstr{I}(x \seqc y) = \abstr{I}(x) \seqc \abstr{I}(y) & \axiom{TI4}  \\
\abstr{I}(\delay(x)) = \delay(\abstr{I}(x))            & \axiom{DRTI} 
\end{axcol}
\qquad
\begin{axcol}
x \parc y =
     (x \leftm y \altc y \leftm x) \altc x \commm y   & \axiom{CM1}   \\
\cts{a} \leftm x = \cts{a} \seqc x                    & \axiom{CM2DR} \\
\cts{a} \seqc x \leftm y = \cts{a} \seqc (x \parc y)  & \axiom{CM3DR} \\
\delay(x) \leftm \timeout(y) = \cts{\dead}            & \axiom{DRCM1} \\
\delay(x) \leftm (\timeout(y) \altc \delay(z)) = \delay(x \leftm z)
                                                      & \axiom{DRCM2} \\
(x \altc y) \leftm z = x \leftm z \altc y \leftm z    & \axiom{CM4}   \\ 
\cts{a} \seqc x \commm \cts{b} = (\cts{a} \commm \cts{b}) \seqc x
                                                      & \axiom{CM5DR} \\
\cts{a} \commm \cts{b} \seqc x = (\cts{a} \commm \cts{b}) \seqc x
                                                      & \axiom{CM6DR} \\
\cts{a} \seqc x \commm \cts{b} \seqc y =
           (\cts{a} \commm \cts{b}) \seqc (x \parc y) & \axiom{CM7DR} \\
\timeout(x) \commm \delay(y) = \cts{\dead}            & \axiom{DRCM3} \\
\delay(x) \commm \timeout(y) = \cts{\dead}            & \axiom{DRCM4} \\
\delay(x) \commm \delay(y) = \delay(x \commm y)       & \axiom{DRCM5} \\
(x \altc y) \commm z = x \commm z \altc y \commm z    & \axiom{CM8}   \\
x \commm (y \altc z) = x \commm y \altc x \commm z    & \axiom{CM9} 
\\ {} \\
\cts{a} \commm \cts{b} = \cts{c}\;\; \mif \commf(a,b) = c 
                                                      & \axiom{CFDR} 
\\ {} \\
\timeout(\cts{a}) = \cts{a}                           & \axiom{DRTO1} \\
\timeout(x \altc y) = \timeout(x) \altc \timeout(y)   & \axiom{DRTO2} \\
\timeout(x \seqc y) = \timeout(x) \seqc y             & \axiom{DRTO3} \\
\timeout(\delay(x)) = \cts{\dead}                     & \axiom{DRTO4} 
\end{axcol}
\\ {} \\
\begin{axcol}
\cts{a} \seqc \cts{\tau} = \cts{a}                    & \axiom{DRB1} \\
\cts{a} \seqc 
(\cts{\tau} \seqc (\timeout(x) \altc y) \altc \timeout(x)) = 
                  \cts{a} \seqc (\timeout(x) \altc y) & \axiom{DRB2} \\
\cts{a} \seqc 
(\cts{\tau} \seqc (\timeout(x) \altc y) \altc y) = 
                  \cts{a} \seqc (\timeout(x) \altc y) & \axiom{DRB3} \\
\cts{a} \seqc 
(\delay(\cts{\tau} \seqc x) \altc \timeout(y)) = 
          \cts{a} \seqc (\delay(x) \altc \timeout(y)) & \axiom{DRB4} 
\end{axcol}
\end{array}
\end{eqntbl}
\vspace*{3ex} \par
\end{table}%
In this table, 
$a$, $b$, and $c$ stand for arbitrary members of $\Acttd$ and
$H$ and $I$ stand for arbitrary subsets of $\Act$.

The axiom names in Table~\ref{axioms-ACPdrtt} have been chosen such that 
it is clear whether an axiom is an axiom from the untimed case without 
any adaptation, an axiom from the untimed case with a minor adaptation 
(with suffix DR), or a completely new axiom (with prefix DR).

\ACP$_\drt^\tau$ is the core of the process algebra \ACP$^\drt_\tau$
presented in~\cite{BM02a}.
In \ACP$_\drt^\tau$, the \emph{deadlocked process} constant $\didead$ 
and the binary \emph{relative initialization} operator $\drinit{}$ from 
\ACP$^\drt_\tau$ have been omitted, the binary \emph{relative delay}
operator $\drdelay{}$ has been replaced by the unary one-time-slice 
delay operator $\delay$, and the binary \emph{relative time-out} 
operator $\drtout{}$ has been replaced by the unary current-time-slice 
time-out operator $\timeout$.
The omitted constant and operator and the replaced operators have 
counterparts that are basic in the case of absolute timing, parametric 
timing or continuous time scale.
However, the omitted constant and operator are not basic in the case of
relative timing and discrete time scale and the replaced operators can 
be defined in terms of their replacements in the case of relative timing 
and discrete time scale.
In~\cite{BM02a}, what is omitted or replaced in \ACP$_\drt^\tau$ is 
relevant to the coherence of the different process algebras presented 
there.

Two subtheories of \ACP$_\drt^\tau$ that will be referred to later are
\BPA$_\drt^\tau$ and \ACP$_\drt$.
\BPA$_\drt^\tau$ is the restriction of \ACP$_\drt^\tau$ 
obtained by omitting the operators $\parc$, $\leftm$, $\commm$, and 
$\encap{H}$ (for $H \subseteq \Act$) and the axioms in which these 
operators occur. 
\ACP$_\drt$ is the restriction of \ACP$_\drt^\tau$ 
obtained by omitting the constant $\cts{\tau}$, the operators 
$\abstr{I}$ (for $I \subseteq \Act$), and the axioms in which this 
constant or these operators occur. 

The notation $\delay^n(t)$, where $n \in \Nat$, is used for the $n$-fold 
application of $\delay$ to $t$, i.e.\ $\delay^0(t) = t$ and 
$\delay^{n+1}(t) = \delay(\delay^{n}(t))$.

The commutativity and associativity of the operators $\altc$ and 
$\parc$ permit the use of the notations 
$\vAltc{i \in \mathcal{I}} t_i$ and $\vParc{i \in {\cal I}} t_i$, 
where $\mathcal{I} = \set{i_1,\ldots,i_n}$,
for the alternative composition $t_{i_1} \altc \ldots \altc t_{i_n}$
and the parallel composition $t_{i_1} \parc \ldots \parc t_{i_n}$,
respec\-tively.
The convention is used that $\vAltc{i \in {\cal I}} t_i$ and
$\vParc{i \in {\cal I}} t_i$ stand for $\cts{\dead}$ if 
$\mathcal{I} = \emptyset$.
If $\cI = \set{i \in \Nat \where \phi_1(i) \Land \ldots \Land \phi_n(i)}$, 
we write $\vAltc{\phi_1(i),\ldots,\phi_n(i)} t_i$ for 
$\vAltc{i \in \cI} t_i$ and $\vParc{\phi_1(i), \ldots, \phi_n(i)} t_i$ 
for $\vParc{i \in \cI} t_i$.

An \ACP$_\drt^\tau$ term $t'$ is said to be a \emph{summand} of an 
\ACP$_\drt^\tau$ term $t$ if there exists a \ACP$_\drt^\tau$ term $t''$
such that $t = t' \altc t''$ or $t = t'$ is derivable from axioms A1 and 
A2.

All processes that can be denoted by a closed \ACP$_\drt^\tau$ term can
be denoted by a basic term over \ACP$_\drt^\tau$.
The set $\cB$ of \emph{basic terms} over \ACP$_\drt^\tau$ is the 
smallest set satisfying:
\begin{itemize}
\item
if $a \in \Acttd$, then $\cts{a} \in \cB$;
\item
if $a \in \Actt$ and $t \in \cB$, then $\cts{a} \seqc t \in \cB$;
\item
if $t \in \cB$, then $\delay(t) \in \cB$;
\item
if $t,t' \in \cB$, then $t \altc t' \in \cB$.
\end{itemize}
Modulo axioms A1 and A2, each basic term from $\cB$ is of the following 
form:
\begin{ldispl}
\Altc{1 \leq i \leq n} \cts{a_i} \seqc t_i \altc
\Altc{1 \leq j \leq m} \cts{b_j} \altc 
\Altc{1 \leq k \leq l} \delay(t'_k)\;,
\end{ldispl}%
where
$n,m,l \in \Nat$,
$a_i \in \Actt$ and $t_i \in \cB$ for $1 \leq i \leq n$,
$b_j \in \Actt$ for $1 \leq j \leq m$, and
$t'_k \in \cB$ for $1 \leq k \leq l$. 
Moreover, modulo axioms A1 and A2, each basic term from~$\cB$ can be 
proved equal to a basic term of this form where $l \leq 1$ by 
applications of axiom DRT1.

The following theorem states that each closed \ACP$_\drt^\tau$ term can 
be reduced to a basic term.
\begin{theorem}[Elimination]
\label{theorem-elim}
For each closed \ACP$_\drt^\tau$ term $t$, there exists a basic term 
$t' \in \cB$ such that $t = t'$ is derivable from the axioms of 
\ACP$_\drt^\tau$.
\end{theorem}
\begin{proof}
The proof goes by induction on the size of $t$.
Corollary~5.3.2.8 from~\cite{Ver97a} states that, for each closed term 
$t$ of \ACP$_\drt$, there exists a basic term $t' \in \cB$ such that 
$t = t'$ is derivable from the axioms of \ACP$_\drt$.%
\footnote{In~\cite{Ver97a}, \ACP$_\drt$ is called \ACP$_\drt^-$-ID.}
The proof of this corollary given in~\cite{Ver97a} carries over to the 
setting with closed terms in which $\cts{\tau}$ may occur.
Because the axioms of \ACP$_\drt$ are included in the axioms of 
\ACP$_\drt^\tau$, this means that the only case that remains to be 
covered is the case where $t$ is of the form $\abstr{I}(t')$. 
By the induction hypothesis, it is sufficient to prove that, for each 
closed term of \ACP$_\drt^\tau$ of the form $\abstr{I}(t')$ with
$t' \in \cB$, there exists a basic term $t'' \in \cB$ such that 
$\abstr{I}(t') = t''$ is derivable from the axioms of \ACP$_\drt^\tau$.
This is easily proved by induction on the structure of $t'$.
\qed
\end{proof}

In Table~\ref{axioms-std-concur}, some equations concerning parallel
composition are given that are derivable for closed terms from the 
axioms of \ACP$_\drt^\tau$ and valid for rooted branching bisimilarity 
as defined in Section~\ref{sect-standard-bisim}.
\begin{table}[!t]
\caption{Axioms of standard concurrency}
\label{axioms-std-concur}
\begin{eqntbl}
\begin{eqncol}
x \parc y = y \parc x
\\
(x \parc y) \parc z = x \parc (y \parc z)
\\
(x \leftm y) \leftm z = x \leftm (y \parc z)
\end{eqncol}
\qquad
\begin{eqncol}
x \commm y = y \commm x
\\
(x \commm y) \commm z = x \commm (y \commm z)
\\
x \commm (y \leftm z) = (x \commm y) \leftm z
\end{eqncol}
\end{eqntbl}
\end{table}%
These equations are called the \emph{axioms of standard concurrency}.

\begin{theorem}[Standard Concurrency]
\label{theorem-standard-concur}
All closed substitution instances of the axioms of standard concurrency 
are derivable from the axioms of \ACP$_\drt^\tau$.
\end{theorem}
\begin{proof}
The proof goes very similar to the proof of Theorem~4.3.3 
from~\cite{BW90}.
Here, by Theorem~\ref{theorem-elim}, it is sufficient to prove the 
theorem only for all closed \ACP$_\drt^\tau$ terms of the form
$\vAltc{1 \leq i \leq n} \cts{a_i} \seqc t_i \altc
 \vAltc{1 \leq j \leq m} \cts{b_j} \altc 
 \vAltc{1 \leq k \leq l} \delay(t'_k)$ where $l \leq 1$.
The possibility of a summand of the form $\delay(t)$ has only a minor 
effect on the complexity of the proof. 
This is the case because the left merge and communication merge of many 
combinations of summands that involve a summand of the form $\delay(t)$ 
are derivably equal to $\cts{\dead}$ by one of the axioms DRCM1--DRCM4.
\qed
\end{proof}

In many applications, communication is synchronous communication 
between two processes.
In those applications, $\commf(\commf(a,b),c) = \dead$ for all 
$a,b,c \in \Act$.
This kind of communication is called \emph{handshaking communication}.
Under the assumption of handshaking communication, the equation 
$\timeout(x \commm y \commm z) = \cts{\dead}$ is derivable for closed 
terms.
This equation is called the \emph{handshaking axiom}.

An important result is the following expansion theorem, which is useful
in the elimination of parallel compositions in terms of \ACP$_\drt^\tau$ 
in the case where all communication is handshaking communication.
\begin{theorem}[Expansion]
\label{theorem-et-drt}
In \ACP$_\drt^\tau$ extended with the axioms of standard concurrency and 
the handshaking axiom, the following equation is derivable for all 
$n > 2$:
\begin{ldispl}
 \begin{geqns}
 x_1 \parc \ldots \parc x_n =
 \Altc{1 \leq i \leq n}
  x_i \leftm
  \Biggl(\Parc{\textstyle \stack{1 \leq j \leq n,}{j \neq i}}
   x_j\Biggr)
  \altc
 \Altc{\textstyle \stack{1 \leq i \leq n,}{i < j \leq n}}
  (x_i \commm x_j) \leftm
  \Biggl(\Parc{\textstyle \stack{1 \leq k \leq n,}{k \neq i, k \neq j}}
   x_k\Biggr)\;.
 \end{geqns}
\end{ldispl}%
\end{theorem}
\begin{proof}
The proof goes the same as the proof of Theorem~4.3.5 from~\cite{BW90}.
\qed
\end{proof}

We will use the standardized terminology and notation for handshaking
communication that were introduced for \ACP\ in \cite{BK86a}.
Processes send, receive and communicate data at \emph{ports}.
If a port is used for communication between two processes, it is called
\emph{internal}.
Otherwise, it is called \emph{external}.
We write:
\begin{itemize}
\item
$s_i(d)$ for the action of sending datum $d$ at port $i$;
\item
$r_i(d)$ for the action of receiving datum $d$ at port $i$;
\item
$c_i(d)$ for the action of communicating datum $d$ at port $i$.
\end{itemize}
The action $c_i(d)$ is the action that is left when $s_i(d)$ and 
$r_i(d)$ are performed synchronously.

\subsection{\ACP$_\drt^\tau$ with Guarded Recursion}
\label{subsect-recursion}

A closed \ACP$_\drt^\tau$ term denotes a process with a finite upper 
bound to the number of actions that it can perform. 
Guarded recursion allows the description of processes without a finite 
upper bound to the number of actions that it can perform.
In this section, we extend \ACP$_\drt^\tau$ with guarded recursion by 
adding constants for solutions of guarded recursive specifications and 
axioms concerning these additional constants.
We write \ACP$_\drt^\tau$\REC\ for the resulting theory.

Let $\Sigma$ be a signature that includes the signature of 
\BPA$_\drt^\tau$, 
let $X$ be a variable from $\cX$, and 
let $t$ be a term over $\Sigma$ in which $X$ occurs.
Then an occurrence of $X$ in $t$ is \emph{guarded} if no abstraction 
operator occurs in $t$ and $t$ has a subterm of the form $a \seqc t'$ 
where $a \in \Act$ or the form $\delay(t')$ and $t'$ contains this 
occurrence of~$X$.
A term $t$ over $\Sigma$ is a \emph{guarded} term over $\Sigma$ if all 
occurrences of variables in $t$ are guarded. 

Let $\Sigma$ be a signature that includes the signature of 
\BPA$_\drt^\tau$.
Then a \emph{recursive specification} over $\Sigma$ is a set 
$\set{X_i = t_i \where i \in I}$, 
where $I$ is an index set, 
each $X_i$ is a variable from $\cX$, 
each $t_i$ is a term over $\Sigma$ in which only variables from 
$\set{X_i \where i \in I}$ occur, 
and $X_i \neq X_j$ for all $i,j \in I$ with $i \neq j$.
We write $\vars(E)$, where $E$ is a recursive specification
$\set{X_i = t_i \where i \in I}$ over $\Sigma$, for the set 
$\set{X_i \where i \in I}$.
Let $E$ be a recursive specification over $\Sigma$ and 
let $X \in \vars(E)$. 
Then the unique equation $X = t\; \in \;E$ is called the 
\emph{recursion equation for $X$ in $E$}.

Let $\Sigma$ be a signature that includes the signature of 
\BPA$_\drt^\tau$, and
let $\Phi$ be a set of equations over the signature $\Sigma$ that 
includes the axioms of \BPA$_\drt^\tau$.
A~recursive specification $\set{X_i = t_i \where i \in I}$ over $\Sigma$ 
is \emph{guarded modulo $\Phi$} if each $t_i$ is rewritable to a guarded 
term over $\Sigma$ by using the equations from $\Phi$ in either 
direction and/or the equations in 
$\set{X_j = t_j \where j \in I \Land i \neq j}$ from left to right.

A solution of a guarded recursive specification 
$\set{X_i = t_i \where i \in I}$ in a semantics of \ACP$_\drt^\tau$\REC\ 
is a set $\set{p_i \where i \in I}$ of processes in that semantics such 
that each equation in $\set{X_i = t_i \where i \in I}$ holds if, for 
each $i \in I$, $X_i$ is assigned $p_i$.
Here, $p_i$ is said to be the $X_i$-component of the solution.
A guarded recursive specification has a unique solution for rooted 
branching bisimilarity as defined on an operational semantics of 
\ACP$_\drt^\tau$\REC\ in Section~\ref{sect-standard-bisim}.
A recursive specification that is not guarded may not have a unique 
solution.
For example, $\set{x = \cts{a} \altc x}$ and $\set{x = x \seqc \cts{a}}$
have infinitely many solutions for rooted branching bisimilarity.

Below, for particular signatures $\Sigma$ that include the signature of 
\BPA$_\drt^\tau$ and particular sets $\Phi$ of equations over $\Sigma$ 
that include the axioms of \BPA$_\drt^\tau$, for each recursive 
specification $E$ over $\Sigma$ that is guarded modulo $\Phi$ and 
$X \in \vars(E)$, a constant $\rec{X}{E}$ that stands for the 
$X$-component of the unique solution of $E$ will be introduced. 
The notation $\rec{t}{E}$ will be used for $t$ with, for all 
$X \in \vars(E)$, all occurrences of $X$ in $t$ replaced by 
$\rec{X}{E}$.

The signature and axiom system of \ACP$_\drt^\tau$\REC\ are 
$\Union_{i \in \Nat} \Sigma_i$ and $\Union_{i \in \Nat} \Phi_i$, 
respectively, where:
\begin{itemize}
\item
$\Sigma_0$ is the signature of \ACP$_\drt^\tau$;
\item
$\Phi_0$ is the axiom system of \ACP$_\drt^\tau$;
\item
$\Sigma_{i+1}$ is $\Sigma_i$ with added, for each recursive 
specification $E$ over $\Sigma_i$ that is guarded modulo $\Phi_i$ and 
$X \in \vars(E)$, a \emph{recursion} constant $\rec{X}{E}$;
\item
$\Phi_{i+1}$ is $\Phi_i$ with added, for each recursive 
specification $E$ over $\Sigma_i$ that is guarded modulo $\Phi_i$ and 
$X \in \vars(E)$, the equation $\rec{X}{E} = \rec{t}{E}$ for the unique 
term $t$ over $\Sigma_i$ such that $X = t\; \in \;E$ and the conditional 
equation $E \Limpl X = \rec{X}{E}$.
\end{itemize}

\sloppy
The equations and conditional equations added to the axiom system of 
\ACP$_\drt^\tau$ to obtain the axiom system of \ACP$_\drt^\tau$\REC\ are
the instances of the axiom schemas RDP and RSP, respectively, given in 
Table~\ref{axioms-REC}.
\begin{table}[!t]
\caption{Additional axioms for \ACP$_\drt^\tau$\REC}
\label{axioms-REC}
\begin{eqntbl}
\begin{saxcol}
\rec{X}{E} = \rec{t}{E}          & \mif X = t\; \in \;E & \axiom{RDP} \\
E \Limpl X = \rec{X}{E}          & \mif X \in \vars(E)  & \axiom{RSP} 
\end{saxcol}
\end{eqntbl}
\end{table}
In these axiom schemas, 
$X$ stands for an arbitrary variable from $\cX$, 
$t$ stands for an arbitrary \ACP$_\drt^\tau$\REC\ term, and
$E$ stands for an arbitrary recursive specification over the signature 
of \ACP$_\drt^\tau$\REC\ that is guarded modulo the axioms of 
ACP$_\drt^\tau$\REC.
Side conditions restrict what $X$, $t$ and $E$ stand for.

For a fixed $E$, the equations $\rec{X}{E} = \rec{t}{E}$ and the 
conditional equations $E \Limpl X \!=\! \rec{X}{E}$ express that the 
constants $\rec{X}{E}$ make up a solution of $E$ and that this solution 
is the only one.

Because conditional equations must be dealt with in 
\ACP$_\drt^\tau$\REC, it is understood that conditional equational logic 
is used in deriving equations from the axioms of \ACP$_\drt^\tau$\REC.
A complete inference system for conditional equational logic can be 
found in~\cite{BW90,Gog21a}.

In the setting of \ACP$_\drt^\tau$\REC, we use $a$, 
where $a \in \Acttd$, as an abbreviation of 
$\rec{X}{X = \cts{a} \altc \delay(X)}$.
For $a \in \Actt$, this means that $a$ denotes the process that performs 
the action $a$ in the current or any future time slice and after that 
immediately terminates successfully.

The abbreviation just introduced can be used in guarded recursive 
specifications because \ACP$_\drt^\tau$\REC\ allows nested recursion.
This shows that nested recursion can be convenient.
However, \ACP$_\drt^\tau$\REC\ would not be less expressive with 
unnested recursion.
\begin{theorem}
\label{theorem-nested-rec}
For each recursion constant $\rec{X}{E}$ of \ACP$_\drt^\tau$\REC, there 
exists a recursion constant $\rec{X}{E'}$ of \ACP$_\drt^\tau$\REC\ 
with no recursion constant of \ACP$_\drt^\tau$\REC\ occurring in $E'$ 
such that $\rec{X}{E} = \rec{X}{E'}$ is derivable from the axioms of 
\ACP$_\drt^\tau$\REC.
\end{theorem}
\begin{proof}
In this proof, $\Sigma_n$ refers to the signature $\Sigma_n$ from the 
definition of the signature of \ACP$_\drt^\tau$\REC.
The proof goes by induction on the nesting level of $\rec{X}{E}$, i.e.\ 
the smallest $n > 0$ such that $\rec{X}{E} \in \Sigma_n$.
The basis step is trivial because there are no occurrences of recursion 
constants of \ACP$_\drt^\tau$\REC\ in $E$ if $\rec{X}{E} \in \Sigma_1$.
The inductive step is proved in the following way.
Assume that all recursion constants occurring in $E$ belong to 
$\Sigma_1$.
This is allowed by the induction hypothesis.
Moreover, assume that, for all recursion constants $\rec{X_1}{E_1}$ and 
$\rec{X_2}{E_2}$ that occur in $E$, 
(a)~$\vars(E_1) \inter \vars(E_2) = \emptyset$ if $E_1 \neq E_2$ and 
(b)~$\vars(E) \inter \vars(E_1) = \emptyset$.
This is allowed by the fact that equality of a recursion constant of 
\ACP$_\drt^\tau$\REC\ and a variable renaming of it is always derivable 
from RDP and RSP.
For $E'$, take the guarded recursive specification obtained from $E$ by
first replacing each recursion constant $\rec{X''}{E''}$ occurring in 
$E$ by the right-hand side of the recursion equation for $X''$ in $E''$ 
and then, for each $\rec{X''}{E''}$ occurring in $E$ adding the 
equations in $E''$ to $E$.
Be aware that 
(a)~for each right-hand side $t$ of an equation in $E$, $\rec{t}{E}$ and 
$\rec{t}{E'}$ stand for the same term and 
(b)~for each right-hand side $t$ of an equation in those $E''$ for which 
there occurs a recursion constant $\rec{X''}{E''}$ in $E$, $\rec{t}{E''}$ 
and $\rec{t}{E'}$ stand for the same term.
Now, first apply RDP for each equation in $E$ and each equation in those 
recursive specifications $E''$ for which there occurs a recursion 
constant $\rec{X''}{E''}$ in $E$ and then apply the instance of RSP for 
$X$ to the resulting set of equations. 
This yields $\rec{X}{E} = \rec{X}{E'}$.
\qed
\end{proof}

To my knowledge, in the setting of \ACP-style process algebras, nested 
guarded recursion has only been considered before in~\cite{Pon91a}.
In that paper, only a restricted form of nested guarded recursion is 
considered.

A system is often described by a term of the form
$\encap{H}(\rec{X_1}{E_1} \parc \ldots \parc \rec{X_n}{E_n})$, i.e.\
as the encapsulated parallel composition of a number of processes that
are described by guarded recursive specifications.
The first step in analyzing such a system is usually the extraction of 
a guarded recursive specification of the system from the term 
describing it.
This involves mere algebraic calculations, in which the expansion
theorem plays an important part, and finally application of RSP.
In case the functional correctness is analyzed, we can proceed by 
extracting a guarded recursive specification from a term of the form 
$\abstr{I}(\rec{X}{E})$ or the form $\drtfp(\abstr{I}(\rec{X}{E}))$,
where $E$ is the result of the first step and $\drtfp$ is the time-free 
projection operator described below.
This involves again application of RSP.

Suppose that $E$ is a guarded recursive specification that includes the 
equation 
\begin{ldispl}
X = 
\vAltc{i \in \cI} \cts{a_i} \seqc t_i \altc
\vAltc{j \in \cJ} \cts{b_j} \altc \delay(X)\;.
\end{ldispl}%
In that case, we can derive the equation
\begin{ldispl}
\rec{X}{E} =
\vAltc{i \in \cI} a_i \seqc \rec{t_i}{E} \altc \vAltc{j \in \cJ} b_j\;.
\end{ldispl}%
We make use of this fact in each of the guarded recursive specifications 
given in Section~\ref{sect-par-protocol}.  

We usually write $X$ for $\rec{X}{E}$ if $E$ is clear from the context.
In those cases, it should also be clear from the context that we use 
$X$ as a constant.

We write \BPA$_\drt^\tau$\REC\ for the theory that is obtained by 
replacing in the definition of the signature and axiom system of
\ACP$_\drt^\tau$\REC\ given above all occurrences of 
``\ACP$_\drt^\tau$'' by ``\BPA$_\drt^\tau$''.
\BPA$_\drt^\tau$\REC\ plays a role in Section~\ref{sect-coarser-eqv}.

In Sections~\ref{sect-standard-bisim} and~\ref{sect-coarser-eqv}, we 
write $\PTerm$ for the set of all closed \ACP$_\drt^\tau$\REC\ terms.

\subsection{Time-free projection}
\label{subsect-other-ext}

\ACP$_\drt^\tau$\REC\ can be extended with the unary 
\emph{time-free projection} operator $\drtfp$.
This operator makes the process denoted by its operand time free. 
For any time-free process, the following holds: it is always capable
of idling till a future time slice and it is never bound to idle till
a future time slice.
Thus, time-free projection amounts to abstraction from the timing of 
actions.
The defining axioms for the time-free projection operator are given in 
Table~\ref{axioms-drtfp}.
\begin{table}[!t]
\caption{Defining axioms of the time-free projection operator}
\label{axioms-drtfp}
\begin{eqntbl}
\begin{axcol}
\drtfp(\cts{a}) = \rec{X}{X = \cts{a} \altc \delay(X)}
                                                    & \axiom{DRTFP1} \\
\drtfp(x \altc y) = \drtfp(x) \altc \drtfp(y)       & \axiom{DRTFP2} \\
\drtfp(x \seqc y) = \drtfp(x) \seqc \drtfp(y)       & \axiom{DRTFP3} \\
\drtfp(\delay(x)) = \drtfp(x)                       & \axiom{DRTFP4}
\end{axcol}
\end{eqntbl}
\end{table}%
In this table, $a$ stands for an arbitrary member of $\Acttd$.

The time-free projection operator relates the timed theory 
\ACP$_\drt^\tau$\REC\ to the untimed theory \ACP$^\tau$\REC\
(\ACP$^\tau$ from~\cite{BW90} with guarded recursion as in 
Section~\ref{subsect-recursion}).
For each \ACP$^\tau$\REC\ term $t$, let $\epsilon(t)$ be the 
\ACP$_\drt^\tau$\REC\ term obtained by replacing, for each 
$a \in \Acttd$, all occurrences of $a$ in $t$ by 
$\rec{X}{X = \cts{a} \altc \delay(X)}$.
Then, for each equation $t = t'$ that can be derived from the axioms of 
\ACP$^\tau$\REC, the equation $\epsilon(t) = \epsilon(t')$ can be 
derived from the axioms of \ACP$_\drt^\tau$\REC. 
Moreover, for each \ACP$_\drt^\tau$\REC\ term $t$, there exists an 
\ACP$^\tau$\REC\ term $t'$ such that $\drtfp(t) = \epsilon(t')$ is 
derivable from the axioms of \ACP$_\drt^\tau$\REC. 

\section{The PAR Protocol}
\label{sect-par-protocol}

In this section, \ACP$_\drt^\tau$\REC\ is used to formally specify a 
simple version of the communication protocol known as the PAR protocol 
and to analyze it.
Timing is essential for the correct behaviour of this protocol.
The treatment of the PAR protocol in this section is largely a slightly
adapted shortened version of its treatment in Sections~2.2.1, 2.2.4, 
and~6.2.3 of~\cite{BM02a}.%
\footnote
{The treatment of the PAR protocol in~\cite{BM02a} have been copied
 almost verbatim without mentioning its origin in at least one other
 publication.}

\subsection{Informal Description of the PAR Protocol}
\label{subsect-informal}

As its name suggests, the PAR protocol is a communication protocol based 
on time-outs of positive acknowledgements.
The sender waits for a positive acknowledgement before a new datum is
transmitted.
If an acknowledgement is not received within a complete protocol cycle,
the old datum is retransmitted.
In order to avoid duplicates due to retransmission, data are labeled
with an alternating bit from $B = \set{0,1}$.
The configuration of the PAR protocol is shown in Fig.~\ref{fig-par}
by means of a connection diagram.
\begin{figure}[!ht]
\centering
\setlength{\unitlength}{.2em}
\begin{picture}(150,50)(0,0)
\put(10,20){\line(1,0){10}}
\put(10,25){\makebox(0,0){1}}
\put(30,20){\circle{20}}
\put(30,20){\makebox(0,0){$S$}}
\put(49,20){\makebox(0,0){$t_S,t_S'$}}
\put(38,25){\line(2,1){22}}
\put(48,35){\makebox(0,0){3}}
\put(75,36){\oval(30,10)}
\put(75,36){\makebox(0,0){$K$}}
\put(75,28){\makebox(0,0){$t_K$}}
\put(90,36){\line(2,-1){22}}
\put(102,35){\makebox(0,0){4}}
\put(38,15){\line(2,-1){22}}
\put(48,5){\makebox(0,0){5}}
\put(75,4){\oval(30,10)}
\put(75,4){\makebox(0,0){$L$}}
\put(75,12){\makebox(0,0){$t_L$}}
\put(90,4){\line(2,1){22}}
\put(102,5){\makebox(0,0){6}}
\put(120,20){\circle{20}}
\put(120,20){\makebox(0,0){$R$}}
\put(101,20){\makebox(0,0){$t_R,t_R'$}}
\put(130,20){\line(1,0){10}}
\put(140,25){\makebox(0,0){2}}
\end{picture}
\par
\caption{Connection diagram for PAR protocol}
\label{fig-par}
\end{figure}

We have a sender process $S$, a receiver process $R$, and two channels
$K$ and $L$.
The process $S$ waits until a datum $d$ is offered at an external port
(port~$1$).
When a datum is offered at this port, $S$ consumes it, packs it with an
alternating bit $b$ in a frame $\tup{d,b}$, and then delivers the frame
at an internal port used for sending (port~$3$).
Next, $S$ waits until an acknowledgement $\nm{ack}$ is offered at an
internal port used for receiving (port~$5$).
When the acknowledgement does not arrive within a certain time period,
$S$ delivers the same frame again and goes back to waiting for an
acknowledgement.
When the acknowledgement arrives within that time period, $S$ goes back
to waiting for a datum, but the alternating bit changes to $(1-b)$.
The process $R$ waits until a frame $\tup{d,b}$ is offered at an 
internal port used for receiving (port~$4$).
When a frame is offered at this port, $R$ consumes it, unpacks it, and
then delivers the datum $d$ at an external port (port~$2$) if the
alternating bit $b$ is the right one and in any case an acknowledgement
$\nm{ack}$ at an internal port used for sending (port~$6$).
After that, $R$ goes back to waiting for a frame, but the right bit
changes to $(1-b)$ if the alternating bit was the right one.
The processes $K$ and $L$ pass on frames from an internal port of $S$
to an internal port of $R$ and acknowledgements from an internal port
of $R$ to an internal port of $S$, respectively.
Because the channels are supposed to be unreliable, they may produce an
error instead of passing on frames or acknowledgements.
The times $t_S,t_R,t_K,t_L$ are the times that it takes the different
processes to pack and deliver, to unpack and deliver or simply to
deliver what they consume.
The time $t_S'$ is the time-out time of the sender, i.e., the time
after which it retransmits a datum in the case where it is still
waiting for an acknowledgement.
The time $t_R'$ is the time that it takes the receiver to produce and
deliver an acknowledgement.

\subsection{Formal Specification of the PAR Protocol}
\label{subsect-specification}

We assume that the times $t_S,t_R,t_K,t_L,t_S',t_R'$ are non-zero.
We also assume a finite set of data $D$.
Let $F = D \x B$ be the set of frames.
For $d \in D$ and $b \in B$, we write $d,b$ for the frame $\tup{d,b}$.
We use the standardized notation for handshaking communication
introduced in Section~\ref{sect-prelims}.
The recursive specification of the sender consists of the following
equations:
\pagebreak[2]
\begin{ldispl}
 \begin{aeqns}
 S & = &
  S_0\;,
 \eqnsep
 S_b & = &
  \Altc{d \in D} \cts{r_1(d)} \seqc \delay^{t_S}(\nm{SF}_{d,b}) \altc
  \delay(S_b)
 \cndsep
 \cnd{(\mathrm{for\; every}\; b \in B),}
 \eqnsep
 \nm{SF}_{d,b} & = &
  \cts{s_3(d,b)} \seqc
   \left(\Altc{k < t_S'}
          \delay^{k}(\cts{r_5(\nm{ack})}) \seqc S_{1 - b} \altc
   \delay^{t_S'}(\nm{SF}_{d,b})\right)
 \cndsep
 \cnd{(\mathrm{for\; every}\; d \in D \;
       \mathrm{and}\; b \in B).}
 \end{aeqns}
\end{ldispl}%
The recursive specification of the receiver consists of the following
equations:
\begin{ldispl}
 \begin{aeqns}
 R & = &
  R_0\;,
 \eqnsep
 R_b & = &
  \Altc{d \in D}
   \cts{r_4(d,b)} \seqc \delay^{t_R}(\cts{s_2(d)}) \seqc
   \delay^{t_R'}(\cts{s_6(\nm{ack})}) \seqc R_{1 - b}
  \\ & &
  {} \altc
  \Altc{d \in D}
   \cts{r_4(d,1-b)} \seqc
                \delay^{t_R'}(\cts{s_6(\nm{ack})}) \seqc R_{b} \altc
  \delay(R_b)
 \cndsep
 \cnd{(\mathrm{for\; every}\; b \in B).}
 \end{aeqns}
\end{ldispl}%
Each of the channels is recursively defined by a single equation:
\begin{ldispl}
 \begin{aeqns}
 K & = &
  \Altc{f \in F}
   \cts{r_3(f)} \seqc
    \left(\delay^{t_K}(\cts{s_4(f)}) \altc
     \Altc{k \leq t_K} \delay^{k}(\cts{\nm{error}})\right) \seqc K
   \altc
  \delay(K)\;,
 \end{aeqns}
\end{ldispl}%
\begin{ldispl}
 \begin{aeqns}
 L & = &
  \cts{r_6(\nm{ack})} \seqc
   \left(\delay^{t_L}(\cts{s_5(\nm{ack})}) \altc
    \Altc{k \leq t_L} \delay^{k}(\cts{\nm{error}})\right) \seqc L
   \altc
  \delay(L)\;.
 \end{aeqns}
\end{ldispl}%
The whole system is described by the following term:
\begin{ldispl}
 \begin{geqns}
 \encap{H}(S \parc K \parc L \parc R)\;,
 \end{geqns}
\end{ldispl}%
where
\begin{ldispl}
\displstretch
 \begin{aeqns}
 H & = &
  \set{s_i(f),r_i(f) \where i \in \set{3,4}, f \in F} \union
  \set{s_i(\nm{ack}),r_i(\nm{ack}) \where i \in \set{5,6}}\;.
 \end{aeqns}
\end{ldispl}%
This protocol is only correct if the time-out time $t_S'$ is longer
than a complete protocol cycle, i.e., if
$t_S' > t_K + t_R + t_R' + t_L$.
If the time-out time is shorter than a complete protocol cycle, the
time-out is called premature.
In that case, while an acknowledgement is still on the way, the sender
will retransmit the current frame.
When the acknowledgement finally arrives, the sender will treat this
acknowledgement as an acknowledgement of the retransmitted frame.
However, an acknowledgement of the retransmitted frame may be on the
way.
If the next frame transmitted gets lost and the latter acknowledgement
arrives, no retransmission of that frame will follow and the protocol
will fail.

\subsection{Analysis of the PAR Protocol: Expansion}
\label{subsect-analysis-1}

We can analyze the PAR protocol described in 
Section~\ref{subsect-specification} using a technique similar to the 
basic one used in the case without timing.

In Section~\ref{subsect-specification}, we first gave guarded recursive 
specifications of the sender process $S$, the receiver process $R$ and 
the channel processes $K$ and $L$, and then described the whole PAR 
protocol by the term $\encap{H}(S \parc K \parc L \parc R)$.
Because all communication is handshaking communication, the expansion
theorem for \ACP$_\drt^\tau$ (see Section~\ref{sect-prelims}) is 
applicable.
By using this expansion theorem and RSP, we are able to give a guarded
recursive specification of the whole PAR protocol.

First, we rewrite the recursive specifications of $S$, $R$, $K$ and
$L$, using their equations and the axioms of \ACP$_\drt$, to ones in a
form that is better suited to expansion.

Secondly, we expand the term $\encap{H}(S \parc K \parc L \parc R)$
by repeated application of the expansion theorem.
Except for the first step, we expand a subterm of the right-hand side 
of a previous equation that has not been expanded before.
After a number of applications of the expansion theorem, the terms on 
the left-hand sides of the equations obtained in this way will include 
all unexpanded terms on the right-hand sides iff 
$t_S' > t_K + t_R + t_R' + t_L$. 
At that point, we have that the terms on the left-hand sides of the 
equations make up a solution of the guarded recursive specification 
obtained by replacing all occurrences of these terms in the equations by 
occurrences of corresponding variables.
Let $X$ be the corresponding variable for 
$\encap{H}(S \parc K \parc L \parc R)$. 
Then we derive from RSP that $\encap{H}(S \parc K \parc L \parc R)$
is the $X$-component of the solution of this guarded recursive 
specification.
The guarded recursive specification concerned can easily be rewritten,
using its equations and the axioms of \ACP$_\drt$, to the following
guarded recursive specification:
\begin{ldispl}
\begin{aeqns}
X & = & X_0\;,
\eqnsep
X_b & = &
\Altc{d \in D} \cts{r_1(d)} \seqc \delay^{t_S}(Y_{d,b}) \altc
\delay(X_b)\;,
\eqnsep
Y_{d,b} & = &
\cts{c_3(d,b)} \seqc
\Biggl(\delay^{t_K}(\cts{c_4(d,b)}) \seqc
 \delay^{t_R}(\cts{s_2(d)}) \seqc
 \delay^{t_R'}(\cts{c_6(\nm{ack})}) \seqc Z_{d,b}
 \\ & & \phantom{\cts{c_3(d,b)} \seqc \Biggl(}
 {} \altc
 \Altc{k \leq t_K}
 \delay^{k}(\cts{\nm{error}}) \seqc \delay^{t_S'-k}(Y_{d,b})\Biggr)\;,
\eqnsep
Z_{d,b} & = &
\delay^{t_L}(\cts{c_5(\nm{ack})}) \seqc X_{1-b} \altc
\Altc{k \leq t_L}
 \delay^{k}(\cts{\nm{error}}) \seqc
 \delay^{t_S' - (t_K + t_R + t_R' + k)}(U_{d,b})\;,
\eqnsep
U_{d,b} & = &
\cts{c_3(d,b)} \seqc
\Biggl(\delay^{t_K}(\cts{c_4(d,b)}) \seqc
 \delay^{t_R'}(\cts{c_6(\nm{ack})}) \seqc V_{d,b}
 \\ & & \phantom{\cts{c_3(d,b)} \seqc \Biggl(}
 {} \altc
 \Altc{k \leq t_K}
 \delay^{k}(\cts{\nm{error}}) \seqc \delay^{t_S'-k}(U_{d,b})\Biggr)\;,
\eqnsep
V_{d,b} & = &
\delay^{t_L}(\cts{c_5(\nm{ack})}) \seqc X_{1-b} \altc
\Altc{k \leq t_L}
 \delay^{k}(\cts{\nm{error}}) \seqc
 \delay^{t_S' - (t_K + t_R' + k)}(U_{d,b})\;.
\end{aeqns}
\end{ldispl}%
From this recursive specification we can conclude informally that, if
we abstract from all actions other than the send and receive actions at
the external ports $1$ and $2$ and in addition from the timing of
actions, the whole PAR proto\-col behaves functionally correct, i.e.\ as 
a buffer with capacity one, provided $t_S' > t_K + t_R + t_R' + t_L$.
Section~\ref{subsect-analysis-2} starts with an outline of how this 
conclusion can be drawn in a formal way.

Details of the analysis outlined in this section can be found in 
Section~2.2.4 of~\cite{BM02a}.

\subsection{Analysis of the PAR Protocol: Abstraction}
\label{subsect-analysis-2}
We want to abstract from all actions other than the send and receive
actions at the external ports $1$ and $2$, i.e.\ from the actions in
the set
\begin{ldispl}
\begin{aeqns}
I & = & 
\set{c_i(d,b) \where i \in \set{3,4}, d \in D, b \in B} \union
\set{c_5(\nm{ack}), c_6(\nm{ack}), \nm{error}}\;.
\end{aeqns}
\end{ldispl}%
We can proceed in different ways. 
First of all, we can focus on functional correctness.
This means that we do not only abstract from the actions in the set
$I$, but also from all timing of actions (using the time free
projection operator from Section~\ref{subsect-other-ext}).
After the abstraction from all timing of actions, we can proceed in a
theory without timing, to wit \ACP$^\tau$\REC\ 
(see Section~\ref{subsect-other-ext}).
Starting from the specification of 
$\encap{H}(S \parc K \parc L \parc R)$ 
at the end of Section~\ref{subsect-analysis-1}, we can easily 
calculate that $\drtfp(\encap{H}(S \parc K \parc L \parc R))$ is the 
$X'$-component of the solution of the guarded recursive specification 
that consists of the following equations:
\begin{ldispl}
\begin{aeqns}
X' & = & X'_0\;,
\eqnsep
X'_b & = &
\Altc{d \in D} r_1(d) \seqc Y'_{d,b}\;,
\eqnsep
Y'_{d,b} & = &
 c_3(d,b) \seqc 
 (c_4(d,b) \seqc s_2(d) \seqc c_6(\nm{ack}) \seqc Z'_{d,b} \altc 
  \nm{error} \seqc Y'_{d,b})\;,
\eqnsep
Z'_{d,b} & = & 
 c_5(\nm{ack}) \seqc X'_{1-b} \altc \nm{error} \seqc U'_{d,b}\;,
\eqnsep 
U'_{d,b} & = &
 c_3(d,b) \seqc 
 (c_4(d,b) \seqc c_6(\nm{ack}) \seqc V'_{d,b} \altc 
  \nm{error} \seqc U'_{d,b})\;,
\eqnsep
V'_{d,b} & = & 
 c_5(\nm{ack}) \seqc X'_{1-b} \altc \nm{error} \seqc U'_{d,b}\;.
\end{aeqns}
\end{ldispl}%
% We see immediately that $Z'_{d,b} = V'_{d,b}$. 
Starting from this specification, we can calculate, using axiom 
$x \seqc (\tau \seqc (y \altc z) \altc z) = x \seqc (y \altc z)$ (B2) 
from \ACP$^\tau$, together with CFAR (Cluster Fair Abstraction Rule) to 
remove cycles of silent steps, that
$\abstr{I}(\drtfp(\encap{H}(S \parc K \parc L \parc R)))$, 
which equals $\drtfp(\abstr{I}(\encap{H}(S \parc K \parc L \parc R)))$,
is the solution of the guarded recursive specification that consists of
the following equation:
\begin{ldispl}
 \begin{aeqns}
 B & = &
  \Altc{d \in D} r_1(d) \seqc s_2(d) \seqc B\;.
 \end{aeqns}
\end{ldispl}%
For more information on the silent step in \ACP$^\tau$, including details 
about CFAR, see~\cite{BW90}.
We have obtained a guarded recursive specification of a buffer with
capacity one.
Thus, we see that the PAR protocol is functionally correct. 
We want to stress that, in order to achieve this result, it was 
necessary to calculate first the time-dependent behavior of the whole
protocol, because the PAR protocol is only correct if the timing
parameters are set correctly. 
A complete verification in process algebra without timing is not 
possible without resorting to artificial tricks such as excluding the 
premature time-out of an acknowledgement by inhibiting a time-out so 
long as it does not lead to deadlock (see e.g.~\cite{Vaa90a}).

Next, we can have a look at the performance properties.
Starting from the specification of 
$\encap{H}(S \parc K \parc L \parc R)$ at the end of
Section~\ref{subsect-analysis-1}, we can easily calculate that
$\abstr{I}(\encap{H}(S \parc K \parc L \parc R))$ is the $X''$-component 
of the solution of the guarded recursive  specification that consists of 
the following equations:
\begin{ldispl}
\begin{aeqns}
X'' & = &
\Altc{d \in D} \cts{r_1(d)} \seqc \delay^{t_S}(Y_d'') \altc
\delay(X'')\;,
\eqnsep
Y_d'' & = &
\delay^{t_K}
 (\cts{\tau} \seqc
  \delay^{t_R}(\cts{s_2(d)} \seqc \delay^{t_R'}(Z''))) \altc
\Altc{k \leq t_K}
 \delay^{k}(\cts{\tau} \seqc \delay^{t_S'-k}(Y_d''))\;,
\eqnsep
Z'' & = &
\delay^{t_L}(\cts{\tau} \seqc X'') \altc
\Altc{k \leq t_L}
 \delay^{k}(\cts{\tau} \seqc
 \delay^{t_S' - (t_K + t_R + t_R' + k)}(U''))\;,
\eqnsep
U'' & = &
\delay^{t_K}(\cts{\tau} \seqc \delay^{t_R'}(V'')) \altc
 \Altc{k \leq t_K}
  \delay^{k}(\cts{\tau} \seqc \delay^{t_S'-k}(U''))\;,
\eqnsep
V'' & = &
\delay^{t_L}(\cts{\tau} \seqc X'') \altc
\Altc{k \leq t_L}
 \delay^{k}(\cts{\tau} \seqc
 \delay^{t_S' - (t_K + t_R' + k)}(U''))\;.
\end{aeqns}
\end{ldispl}%
Not many simplifications have been achieved.
This is mainly because we cannot leave out silent steps that occur in
between delays.
In effect, all internal choices that may be made, e.g. whether or not a 
channel forwards a datum correctly, remain visible.
Some initial observations concerning this matter, as well as an 
analysis of a slightly different version of the PAR protocol, were made
in an unpublished paper of Dragan Bo{\u{s}}na{\u{c}}ki (1997).
In any case, from this specification, we can conclude informally that
the protocol takes at least $t_S + t_K + t_R$ time slices between
consumption and delivery of a datum, and in general, between
consumption and delivery we have $t_S + t_K + t_R + i \cdot t_S'$ 
time slices, where $i \geq 0$.
After delivery, at least $t_R' + t_L$ time slices must pass before the
next datum can be consumed, and in general, between delivery and 
readiness for the next consumption, we have $t_R' + t_L$ or
$t_R' + t_L + j \cdot t_S' - t_R$ time slices, where $j > 0$.

Details of the analysis outlined in this section can be found in 
Section~6.2.3 of~\cite{BM02a}.

\section{The Standard Version of Branching Bisimilarity}
\label{sect-standard-bisim}

For the relevant versions of \ACP, the accepted definition of the 
standard version of branching bisimilarity for processes with discrete 
relative timing refers to a two-phase operational semantics of the 
version concerned, i.e.\ an operational semantics where processes can 
make transitions by performing an action in the current time slice or by 
idling till the next time slice.
In Section~\ref{sect-coarser-eqv}, a definition of a variant of the 
standard version of branching bisimilarity for processes with discrete 
relative timing will be given.
That definition refers to a time-stamped operational semantics of 
\ACP$_\drt^\tau$\REC, i.e.\ an operational semantics where processes can 
make transitions by idling till a time slice in which an action can be 
performed and subsequently performing that action.

In this section, a two-phase operational semantics and a time-stamped 
operational semantics of \ACP$_\drt^\tau$\REC\ is presented and the 
standard version of branching bisimilarity for processes with discrete 
relative timing is defined once referring to the two-phase operational 
semantics and once referring to the time-stamped operational semantics.
This way, it becomes easier to see the essential difference between the
standard version of branching bisimilarity for processes with discrete 
relative timing and the variant introduced in 
Section~\ref{sect-coarser-eqv}.

\subsection{A Definition Referring to a Two-Phase Semantics}
\label{subsect-standard-two-phase}

The definition of the standard version of branching bisimilarity given 
below refers to a two-phase structural operational semantics of 
\ACP$_\drt^\tau$\REC\ that consists of:
\begin{itemize}
\item[]
a binary \emph{transition} relation
$\trans{\ph}{\alpha}{\ph} \subseteq \PTerm \x (\PTerm \union \set{\surd})$
for each $\alpha \in \Actt \union \set{\sigma}$, where, for each 
$t \in \PTerm$, not $\trans{t}{\sigma}{\surd}$.
\end{itemize}
The relations from this structural operational semantics describe what 
the processes denoted by terms from $\PTerm$ are capable of doing as 
follows:
\begin{enumerate}
\item[]
if $\alpha \in \Actt$, then $\trans{t}{\alpha}{t'}$ indicates that the 
process denoted by $t$ has the potential to make a transition to the 
process denoted by $t'$ by performing action $\alpha$ in the current 
time slice;
\item[]
if $\alpha = \sigma$, then $\trans{t}{\alpha}{t'}$ indicates that the 
process denoted by $t$ has the potential to make a transition
to the process denoted by $t'$ by idling till the next time slice.
\end{enumerate}
Moreover, $\surd$ denotes the process that is only capable of 
terminating successfully in the current time slice.
There exists no \ACP$_\drt^\tau$\REC\ term that denotes this process.

The relations from the two-phase structural operational semantics of 
\ACP$_\drt^\tau$\REC\ are defined below by means of rules of which some
have negative premises, i.e.\ premises of the form 
$\Lnot\, (\trans{t}{\alpha}{t'})$.
Because of this, it is not so obvious what relations are defined by the 
rules concerned.
An informal explanation is given here.
The relations defined by these rules are the unique ones that constitute
an operational semantics such that $\trans{t}{\alpha}{t'}$ holds in that
operational semantics iff $\trans{t}{\alpha}{t'}$ can be deduced from 
the rules under some set of negative assumptions that hold in that 
operational semantics.
% for all $\alpha \in \Actt \union \set{\sigma}$  
% and $t, t'\in \PTerm \union \set{\surd}$ 
The uniqueness follows from the fact that the rules are such that no
closed substitution instance of the conclusion of a rule depends 
negatively on itself.
Formal details and further explanation can for example be found 
in~\cite{Gla04a}.

We write $\ntrans{t}{\sigma}$ for the set of all premises 
$\Lnot\, (\trans{t}{\sigma}{t'})$ where $t' \in \PTerm$.

The relations from the two-phase structural operational semantics of 
\ACP$_\drt^\tau$\REC\ are the relations defined by the rules given in 
Table~\ref{rules-ACPdrttaurec} as explained above.
\begin{table}[!p]
\caption{Transition rules for \ACP$_\drt^\tau$\REC}
\label{rules-ACPdrttaurec}
\begin{ruletbl}
\Rule
{\phantom{\trans{x}{a}{x'}}}
{\trans{\cts{a}}{a}{\surd}}
\\
\Rule
{\trans{x}{a}{x'}}
{\trans{x \altc y}{a}{x'}}
\quad
\Rule
{\trans{y}{a}{y'}}
{\trans{x \altc y}{a}{y'}}
\quad
\Rule
{\trans{x}{a}{\surd}}
{\trans{x \altc y}{a}{\surd}}
\quad
\Rule
{\trans{y}{a}{\surd}}
{\trans{x \altc y}{a}{\surd}}
\\
\Rule
{\trans{x}{\sigma}{x'},\; \ntrans{y}{\sigma}}
{\trans{x \altc y}{\sigma}{x'}}
\quad
\Rule
{\ntrans{x}{\sigma},\; \trans{y}{\sigma}{y'}}
{\trans{x \altc y}{\sigma}{y'}}
\quad
\Rule
{\trans{x}{\sigma}{x'},\; \trans{y}{\sigma}{y'}}
{\trans{x \altc y}{\sigma}{x' \altc y'}}
\\
\Rule
{\trans{x}{a}{x'}}
{\trans{x \seqc y}{a}{x' \seqc y}}
\quad
\Rule
{\trans{x}{a}{\surd}}
{\trans{x \seqc y}{a}{y}}
\quad
\Rule
{\trans{x}{\sigma}{x'}}
{\trans{x \seqc y}{\sigma}{x' \seqc y}}
\\
\Rule
{\phantom{\trans{x}{a}{x'}}}
{\trans{\delay(x)}{\sigma}{x}}
\\
\Rule
{\trans{x}{a}{x'}}
{\trans{x \parc y}{a}{x' \parc y}}
\quad
\Rule
{\trans{y}{a}{y'}}
{\trans{x \parc y}{a}{x \parc y'}}
\quad
\Rule
{\trans{x}{a}{\surd}}
{\trans{x \parc y}{a}{y}}
\quad
\Rule
{\trans{y}{a}{\surd}}
{\trans{x \parc y}{a}{x}}
\\
\RuleC
{\trans{x}{a}{x'},\; \trans{y}{b}{y'}}
{\trans{x \parc y}{c}{x' \parc y'}}
{\commf(a,b) = c}
\quad
\RuleC
{\trans{x}{a}{x'},\; \trans{y}{b}{\surd}}
{\trans{x \parc y}{c}{x'}}
{\commf(a,b) = c}
\\
\RuleC
{\trans{x}{a}{\surd},\; \trans{y}{b}{y'}}
{\trans{x \parc y}{c}{y'}}
{\commf(a,b) = c}
\quad
\RuleC
{\trans{x}{a}{\surd},\; \trans{y}{b}{\surd}}
{\trans{x \parc y}{c}{\surd}}
{\commf(a,b) = c}
\quad
\Rule
{\trans{x}{\sigma}{x'},\; \trans{y}{\sigma}{y'}}
{\trans{x \parc y}{\sigma}{x' \parc y'}}
\\
\Rule
{\trans{x}{a}{x'}}
{\trans{x \leftm y}{a}{x' \parc y}}
\quad
\Rule
{\trans{x}{a}{\surd}}
{\trans{x \leftm y}{a}{y}}
\quad
\Rule
{\trans{x}{\sigma}{x'},\; \trans{y}{\sigma}{y'}}
{\trans{x \leftm y}{\sigma}{x' \leftm y'}}
\\
\RuleC
{\trans{x}{a}{x'},\; \trans{y}{b}{y'}}
{\trans{x \commm y}{c}{x' \parc y'}}
{\commf(a,b) = c}
\quad
\RuleC
{\trans{x}{a}{x'},\; \trans{y}{b}{\surd}}
{\trans{x \commm y}{c}{x'}}
{\commf(a,b) = c}
\\
\RuleC
{\trans{x}{a}{\surd},\; \trans{y}{b}{y'}}
{\trans{x \commm y}{c}{y'}}
{\commf(a,b) = c}
\quad
\RuleC
{\trans{x}{a}{\surd},\; \trans{y}{b}{\surd}}
{\trans{x \commm y}{c}{\surd}}
{\commf(a,b) = c}
\quad
\Rule
{\trans{x}{\sigma}{x'},\; \trans{y}{\sigma}{y'}}
{\trans{x \commm y}{\sigma}{x' \commm y'}}
\\
\RuleC
{\trans{x}{a}{x'}}
{\trans{\encap{H}(x)}{a}{\encap{H}(x')}}
{a \not\in H}
\quad
\RuleC
{\trans{x}{a}{\surd}}
{\trans{\encap{H}(x)}{a}{\surd}}
{a \not\in H}
\quad
\Rule
{\trans{x}{\sigma}{x'}}
{\trans{\encap{H}(x)}{\sigma}{\encap{H}(x')}}
\\
\RuleC
{\trans{x}{a}{x'}}
{\trans{\abstr{I}(x)}{a}{\abstr{I}(x')}}
{a \not\in I}
\quad
\RuleC
{\trans{x}{a}{\surd}}
{\trans{\abstr{I}(x)}{a}{\surd}}
{a \not\in I}
\\
\RuleC
{\trans{x}{a}{x'}}
{\trans{\abstr{I}(x)}{\tau}{\abstr{I}(x')}}
{a \in I}
\quad
\RuleC
{\trans{x}{a}{\surd}}
{\trans{\abstr{I}(x)}{\tau}{\surd}}
{a \in I}
\quad
\Rule
{\trans{x}{\sigma}{x'}}
{\trans{\abstr{I}(x)}{\sigma}{\abstr{I}(x')}}
\\
\Rule
{\trans{x}{a}{x'}}
{\trans{\timeout(x)}{a}{x'}}
\quad
\Rule
{\trans{x}{a}{\surd}}
{\trans{\timeout(x)}{a}{\surd}}
\\
\RuleC
{\trans{\rec{t}{E}}{a}{x'}}
{\trans{\rec{X}{E}}{a}{x'}}
{X \!\!=\! t \,\in\, E}
\quad
\RuleC
{\trans{\rec{t}{E}}{a}{\surd}}
{\trans{\rec{X}{E}}{a}{\surd}}
{X \!\!=\! t \,\in\, E}
\quad
\RuleC
{\trans{\rec{t}{E}}{\sigma}{x'}}
{\trans{\rec{X}{E}}{\sigma}{x'}}
{X \!\!=\! t \,\in\, E}
\end{ruletbl}
\end{table}%
In this table, 
$a$, $b$, and $c$ stand for arbitrary actions from $\Actt$,
$H$ and $I$ stand for arbitrary subsets of $\Act$,
$X$ stands for an arbitrary variable from $\cX$, 
$t$ stands for an arbitrary \ACP$_\drt^\tau$\REC\ term, and
$E$ stands for an arbitrary guarded recursive specification over 
\ACP$_\drt^\tau$\REC.

In Figure~\ref{fig-two-phase}, the processes that are denoted by the
closed terms
$\delay(\cts{a}) \altc \delay(\cts{b})$ and 
$\cts{\tau} \seqc \delay(\cts{a}) \altc \delay(\cts{b})$ 
according to the two-phase operational semantics are presented 
graphically.
These graphical presentations show the effect of the peculiar transition 
rules for alternative composition with respect to idling till the next 
time slice: in the case of an alternative composition of two processes, 
the choice between the two is resolved at the instant that one of them 
performs its first action, and not before.
\begin{figure}[!h]
\centering
\begin{tsfig}
\begin{array}[t]{c}
\delay(\cts{a}) \altc \delay(\cts{b}) \\
\stepdl{\sigma}\\
\cts{a} \altc \cts{b} \\
\stepsw{a} \stepse{b} \\
\surd \phantom{\stepsw{} \stepse{}} \surd
\end{array}
\qquad \qquad
\begin{array}[t]{c}
\cts{\tau} \seqc \delay(\cts{a}) \altc \delay(\cts{b}) \\
\stepsw{\tau} \stepse{\sigma} \\
\begin{array}[t]{c@{}}
\delay(\cts{a})  \\
\stepdl{\sigma} \\
\cts{a} \\
\stepdl{a} \\
\surd
\end{array}
\phantom{\stepsw{} \! \stepse{}}
\begin{array}[t]{@{}c}
\cts{b}  \\
\stepdr{b} \\
\surd
\end{array}
\end{array}
\end{tsfig}
\caption{Graphical two-phase presentation of two processes}
\label{fig-two-phase}
\end{figure}

Two processes are considered equal if they can simulate each other 
insofar as their observable potentials to make transitions are 
concerned.
This can be dealt with by means of the notion of branching bisimilarity 
introduced in~\cite{GW96a} adapted to processes with discrete relative
timing in the style of~\cite{BB95b}.
 
In the definition of rooted branching bisimilarity below, we denote the 
reflexive and transitive closure of $\step{\sigma}$ by $\midling$.
So $t \midling t'$ indicates that the process denoted by $t'$ is 
reachable from the process denoted by $t$ by idling only. 

A \emph{branching bisimulation} is a binary relation $R$ on 
$\PTerm \union \set{\surd}$ such that, for all $t_1,t_2 \in \PTerm$ with 
$R(t_1,t_2)$, the following transfer conditions hold:
\begin{itemize}
\item 
if $\trans{t_1}{\alpha}{t_1'}$, then  
\begin{itemize}
\item 
either $\alpha \equiv \tau$ and $R(t_1',t_2)$ 
\item
or, for some $n \in \Nat$, 
there exist 
$t^*_0,\ldots,t^*_n \in \PTerm$ and $t_2' \in \PTerm \union \set{\surd}$ 
with $t^*_0 \equiv t_2$ such that 
\begin{itemize}
\item 
for all $i \in \Nat$ with $i < n$, $\trans{t^*_i}{\tau}{t^*_{i+1}}$ and
$R(t_1,t^*_{i+1})$,
\item
$\trans{t^*_n}{\alpha}{t_2'}$, and $R(t_1',t_2')$;
\end{itemize}
\end{itemize}
\item 
if $\trans{t_2}{\alpha}{t_2'}$, then
\begin{itemize}
\item 
either $\alpha \equiv \tau$ and $R(t_1,t_2')$ 
\item
or, for some $n \in \Nat$, 
there exist 
$t^*_0,\ldots,t^*_n \in \PTerm$ and $t_1' \in \PTerm \union \set{\surd}$ 
with $t^*_0 \equiv t_1$ such that 
\begin{itemize}
\item 
for all $i \in \Nat$ with $i < n$, $\trans{t^*_i}{\tau}{t^*_{i+1}}$ and
$R(t^*_{i+1},t_2)$,
\item
$\trans{t^*_n}{\alpha}{t_1'}$, and $R(t_1',t_2')$.
\end{itemize}
\end{itemize}
\end{itemize}
\pagebreak[2]
Two terms $t_1,t_2 \in \PTerm$ are \emph{branching bisimilar}, 
written $t_1 \bisim_\br t_2$, if there exists a branching bisimulation 
$R$ with $R(t_1,t_2)$.
Branching bisimilarity is also called 
\emph{branching bisimulation equivalence}.

Branching bisimilarity is not a congruence with respect to the operators
$\altc$, $\leftm$, and $\commm$. 
This can be remedied by means of a root condition.

If $R$ is a binary relation on $\PTerm \union \set{\surd}$, then a pair 
$(t_1,t_2) \in \PTerm \x \PTerm$ is said to satisfy the 
\emph{root condition} in $R$ if the following condition holds:
\begin{itemize}
\item 
if $\trans{t_1}{\alpha}{t_1'}$, then 
there exists a $t_2' \in \PTerm \union \set{\surd}$ such that 
\smash{$\trans{t_2}{\alpha}{t_2'}$} and $R(t_1',t_2')$;
\item 
if $\trans{t_2}{\alpha}{t_2'}$, then 
there exists a $t_1' \in \PTerm \union \set{\surd}$ such that 
\smash{$\trans{t_1}{\alpha}{t_1'}$} and $R(t_1',t_2')$.
\end{itemize}
Two terms $t_1,t_2 \in \PTerm$ are \emph{rooted branching bisimilar}, 
written $t_1 \bisim_\rb t_2$, if there exists a branching bisimulation 
$R$ with $R(t_1,t_2)$ such that, for all terms $t_1',t_2' \in \PTerm$
with $t_1 \midling t_1'$, $t_2 \midling t_2'$, and $R(t_1',t_2')$, the 
pair $(t_1',t_2')$ satisfies the root condition in $R$.
Rooted branching bisimilarity is also called 
\emph{rooted branching bisimulation equivalence}.

Let $R$ be a branching bisimulation such that $R(t_1,t_2)$ and, 
for all terms $t_1',t_2' \in \PTerm$ with  $t_1 \midling t_1'$, 
$t_2 \midling t_2'$, and $R(t_1',t_2')$, the pair $(t_1',t_2')$ 
satisfies the root condition in $R$.
Then we say that $R$ is a branching bisimulation \emph{witnessing}
$t_1 \bisim_\rb t_2$.

If $R$ is a branching bisimulation witnessing $t_1 \bisim_\rb t_2$, then
not only the pair $(t_1,t_2)$ must satisfy the root condition in $R$, 
but also all other pairs $(t_1',t_2')$ for which $t_1 \midling t_1'$, 
$t_2 \midling t_2'$, and $R(t_1',t_2')$.
Without this additional requirement (relative to branching bisimilarity 
as defined in~\cite{GW96a}), branching bisimilarity is doomed not to be
a congruence with respect to alternative composition, left merge and 
communication merge in the current setting 
(see Theorem~\ref{theorem-rbbisim-congr} below).

\subsection{Properties of Rooted  Branching Bisimilarity}
\label{subsect-sound-complete}

Branching bisimilarity and rooted branching bisimilarity are equivalence 
relations.
\begin{theorem}[Equivalence]
\label{theorem-rbbisim-equiv}
The binary relations $\bisim_\br$ and $\bisim_\rb$ on $\PTerm$ are 
equivalence relations.
\end{theorem}
\begin{proof}
By Lemma~2.5 from~\cite{GW96a}, known as the stuttering lemma, 
$\bisim_\br$ is an instance of the standard notion of branching 
bisimilarity. 
By Proposition~2.11 from~\cite{GW96a}, $\bisim_\br$ is therefore an 
equivalence relation.
The proofs of reflexivity and symmetry of $\bisim_\rb$ go the same as 
the proofs of reflexivity and symmetry of $\bisim_\br$.
To prove that $t \bisim_\rb t'$ and $t' \bisim_\rb t''$ implies  
$t \bisim_\rb t''$, consider the relation composition of witnessing 
branching bisimulations $R$ and $R'$ for $t \bisim_\rb t'$ and 
$t' \bisim_\rb t''$, respectively.
Take arbitrary $t_1,t_2,t_3 \in \PTerm$ such that $R(t_1,t_2)$ and
$R'(t_2,t_3)$.
Two cases have to be distinguished:
(a)~the case where $\tup{t_1,t_2}$ has to satisfy the root condition in 
$R$ and
(b)~the case where $\tup{t_1,t_2}$ does not have to satisfy the root 
condition in $R$. 
If $\tup{t_1,t_2}$ has to satisfy the root condition in $R$, then 
$\tup{t_2,t_3}$ has to satisfy the root condition in $R'$ as well.
Case~(a) follows immediately from the satisfaction of the the root 
condition in $R$ and $R'$ by $\tup{t_1,t_2}$ and $\tup{t_2,t_3}$, 
respectively.
The proof of case~(b) goes the same as in the proof of Proposition~2.11 
from~\cite{GW96a} (for the details, look at the proof of Proposition~7
from~\cite{Bas96a}).
\qed
\end{proof}

Rooted branching bisimilarity is a congruence relation with respect to
all operators from the signature of \ACP$_\drt^\tau$\REC.
\begin{theorem}[Congruence]
\label{theorem-rbbisim-congr}
The binary relation $\bisim_\rb$ on $\PTerm$ is a congruence relation 
with respect to all operators from the signature of 
\ACP$_\drt^\tau$\REC.
\end{theorem}
\begin{proof}
In the proof, the auxiliary notions of weakly rooted branching 
bisimilarity and strong root condition are used.
Two terms $t_1,t_2 \in \PTerm$ are \emph{weakly rooted branching 
bisimilar}, written $t_1 \bisim_\wrb t_2$, if there exists a branching 
bisimulation $R$ with $R(t_1,t_2)$ such that the pair $(t_1,t_2)$ 
satisfies the root condition in~$R$;
if $R$ is a binary relation on $\PTerm \union \set{\surd}$, then a pair 
$(t_1,t_2) \in \PTerm \x \PTerm$ is said to satisfy the 
\emph{strong root condition} in $R$ if, 
for all terms $t_1',t_2' \in \PTerm$ with $t_1 \midling t_1'$, 
$t_2 \midling t_2'$, and $R(t_1',t_2')$, the pair $(t_1',t_2')$ 
satisfies the root condition in~$R$.

It follows directly from the above definitions that $\bisim_\rb$ is a 
congruence with respect to all operators from the signature of 
\ACP$_\drt^\tau$\REC\ iff $\bisim_\wrb$ is a congruence with respect to 
all operators from the signature of \ACP$_\drt^\tau$\REC\ and 
satisfaction of the strong root condition is preserved by all operators 
from the signature of \ACP$_\drt^\tau$\REC. 
It is not hard to see that satisfaction of the strong root condition is 
preserved by all operators from the signature of \ACP$_\drt^\tau$\REC.

The RBB safe congruence format from~\cite{Fok00a} is applicable to 
$\bisim_\wrb$.
Except for the rules for alternative composition, left merge, and 
communication merge concerning the relation $\step{\sigma}$, all rules 
given in  Table~\ref{rules-ACPdrttaurec} satisfy the restrictions of the 
RBB safe format.
However, the preservation of the strong root condition by alternative 
composition, left merge, and communication merge exclude the possibility
that these rules are the cause of non-congruence.
\qed
\end{proof}

Below, the soundness of the axiom system of \ACP$_\drt^\tau$\REC\ with 
respect to~$\bisim_\rb$ for equations between terms from $\PTerm$ will 
be established.

In this paper, the following terminology will be used in soundness proofs: 
(a)~an equation $\eqn$ over a signature $\Sigma$ is said to be 
\emph{valid with respect to} a congruence relation $\simeq$ on the set 
of all closed terms over a signature $\Sigma$ if, for each closed 
substitution instance $t = t'$ of $\eqn$, $t \simeq t'$ and
(b)~a conditional equation $\ceqn$ over a signature $\Sigma$ is said to 
be \emph{valid with respect to} a congruence relation $\simeq$ on the 
set of all closed terms over a signature $\Sigma$ if, for each closed 
substitution instance $\set{t_i = t'_i \where i \in I} \Limpl t = t'$ of 
$\ceqn$, $t \simeq t'$ if $t_i \simeq t'_i$ for each $i \in I$.

\begin{theorem}[Soundness]
\label{theorem-soundness}
For all terms $t, t' \in \PTerm$, $t = t'$ is derivable from the axioms 
of \ACP$_\drt^\tau$\REC\ only if $t \bisim_\rb t'$.
\end{theorem}
\begin{proof}
Because $\bisim_\rb$ is a congruence with respect to all operators from 
the signature of \ACP$_\drt^\tau$\REC, only the validity of each axiom 
of \ACP$_\drt^\tau$\REC\ has to be proved.
The proofs of the validity of the axioms of \ACP$_\drt$ with respect to 
$\bisim$ given in the proof of Theorem~5.2.3.10 from~\cite{Ver97a} carry 
over to the setting with closed terms in which $\cts{\tau}$ may occur.
Because ${\bisim} \subseteq {\bisim_\rb}$, validity with respect to
$\bisim$ implies validity with respect to $\bisim_\rb$.
Therefore, it is sufficient to prove the validity of axioms TI1DR, 
TI2DR, TI3, TI4, DRTI, DRB1--DRB4, RDP, and RSP with respect to 
$\bisim_\rb$.

Below, we write $\csi(\eqn)$, where $\eqn$ is an equation between two
\ACP$_\drt^\tau$\REC\ terms, for the set of all closed substitution
instances of $\eqn$.
Moreover, we write $R_\id$ for the identity relation on 
$\PTerm \union \set{\surd}$.

For each axiom $\ax$ from the above-mentioned axioms, a branching 
bisimulation $R_\ax$ witnessing the validity of $\ax$ can be constructed 
as follows:
\begin{itemize}
\item
if $\ax$ is an instance of one of the axiom schemas TI1DR, TI2DR, TI3, 
TI4, DRTI, RDP or axiom DRB1:
\begin{ldispl}
R_\ax = \set{\tup{t,t'} \where t = t' \in \csi(\ax)} \union R_\id\;;
\end{ldispl}%
\item
if $\ax$ is an instance of axiom schema DRB2:
\begin{ldispl}
R_\ax = 
\set{\tup{t,t'} \where t = t' \in \csi(\ax)} 
\\ \phantom{R_\ax = {}}\,
 {} \union
 \set{\tup{t,t'} \where 
      t = t' \in 
      \csi(\cts{\tau} \seqc (\timeout(x) \altc y) \altc \timeout(x) = 
           \timeout(x) \altc y)}
\union R_\id\;;
\end{ldispl}%
\item
if $\ax$ is an instance of axiom schema DRB3: similar;
\item
if $\ax$ is an instance of axiom schema DRB4: 
\begin{ldispl}
R_\ax = 
\set{\tup{t,t'} \where t = t' \in \csi(\ax)} 
\\ \phantom{R_\ax = {}}\,
 {} \union
 \set{\tup{t,t'} \where 
      t = t' \in 
      \csi(\delay(\cts{\tau} \seqc x) \altc \timeout(y) = 
           \delay(x) \altc \timeout(y))} 
\\ \phantom{R_\ax = {}}\, 
 {}  \union 
 \set{\tup{t,t'} \where t = t' \in \csi(\cts{\tau} \seqc x = x)}
\union R_\id\;;
\end{ldispl}%
\item
if $\ax$ is an instance 
$\set{X_i = t_i \where i \in I} \Limpl 
 X_j = \rec{X_j}{\set{X_i = t_i \where i \in I}}$ 
($j \in I$) of RSP:
\begin{ldispl}
R_\ax 
\\ \; {} = 
\set{\tup{\theta(X_j),\rec{X_j}{\set{X_i = t_i \where i \in I}}} \where 
     j \in I \Land \theta \in \Theta \Land
     \LAND_{i \in I} \theta(X_i) \bisim \theta(t_i)}
\\ \phantom{\; {} = {}}\, 
 {}  \union R_\id\;,
\end{ldispl}%
where 
$\Theta$ is the set of all functions from $\cX$ to $\PTerm$ and 
$\theta(t)$, where $\theta \in \Theta$ and $t \in \PTerm$, stands
for $t$ with, for all $X \in \cX$, all occurrences of $X$ replaced by 
$\theta(X)$.
\end{itemize}
For each equational axiom $\ax$, it is straightforward to check that 
the constructed relation $R_\ax$ is a branching bisimulation witnessing, 
for each closed substitution instance $t = t'$ of $\ax$, 
$t \bisim_\rb t'$.
For each conditional equational axiom $\ax$, i.e.\ for each instance of 
RSP, it is straightforward to check that the constructed relation 
$R_\ax$ is a branching bisimulation witnessing, for each closed 
substitution instance $\set{t_i = t'_i \where i \in I} \Limpl t = t'$ of 
$\ax$, $t \bisim_\rb t'$ if $t_i \bisim_\rb t'_i$ for each $i \in I$.
\qed
\end{proof}

The axiom system of \ACP$_\drt^\tau$\REC\ is complete with respect to 
$\bisim_\rb$ for equations between terms from $\PTerm$ in which only the 
constants and operators of \ACP$_\drt$ occur.
\begin{theorem}[Completeness]
\label{theorem-completeness}
For all terms $t, t' \in \PTerm$ that are closed terms of \ACP$_\drt$, 
$t = t'$ is derivable from the axioms of \ACP$_\drt^\tau$\REC\ if 
$t \bisim_\rb t'$.
\end{theorem}
\begin{proof}
This is Theorem~5.2.3.13 from~\cite{Ver97a}.
\qed
\end{proof}

It is an open problem whether the axiom system of \ACP$_\drt^\tau$\REC\ 
is complete with respect to $\bisim_\rb$ for equations between terms 
from $\PTerm$ in which only the constants and operators of 
\ACP$_\drt^\tau$ occur.

Even if the axiom system of \ACP$_\drt^\tau$\REC\ is complete with 
respect to $\bisim_\rb$ for equations between terms from $\PTerm$ in 
which only the constants and operators of \ACP$_\drt^\tau$ occur, it is 
not complete with respect to $\bisim_\rb$ for equations between terms 
from $\PTerm$.
Consider, for example, 
the \ACP$_\drt^\tau$\REC\ term $\abstr{\set{a,b}}(\rec{X}{E})$, where 
$E = \set{X = \cts{a} \seqc Y, Y = \cts{b} \seqc X \altc \cts{c}}$.
Then $\abstr{\set{a,b}}(\rec{X}{E}) \bisim_\rb \cts{\tau} \seqc \cts{c}$, 
but the equation
$\abstr{\set{a,b}}(\rec{X}{E}) = \cts{\tau} \seqc \cts{c}$
is not derivable from the axioms of \ACP$_\drt^\tau$\REC. 
It is an open problem which additional axioms and/or restrictions on the
form of guarded recursive specifications are needed to yield 
completeness with respect to $\bisim_\rb$ for equations between terms 
from $\PTerm$.

\subsection{A Definition Referring to a Time-Stamped Semantics}
\label{subsect-standard-time-stamped}

The definition of the standard version of branching bisimilarity given 
below refers to a time-stamped structural operational semantics of 
\ACP$_\drt^\tau$\REC\ that consists of:
\begin{itemize}
\item
a binary \emph{transition} relation
$\trans{\ph}{a[n]}{\ph} \subseteq \PTerm \x (\PTerm \union \set{\surd})$
for each $a \in \Actt$ and $n \in \Nat$;
\item
a unary \emph{idling} relation
${\idling{\ph}{n}} \subseteq \PTerm$ for each $n \in \Nat$.
\end{itemize}
The relations from this structural operational semantics describe what 
the processes denoted by terms from $\PTerm$ are capable of doing as 
follows:
\begin{itemize}
\item
$\trans{t}{a[n]}{t'}$ indicates that the process denoted by $t$ has the 
potential to make a transition to the process denoted by $t'$ by 
performing action $a$ in the $n$th-next time slice;
\item
$\idling{t}{n}$ indicates that the process denoted by $t$ is capable of 
idling till the $n$th-next time slice.
\end{itemize}

To define the time-stamped structural operational semantics of 
\ACP$_\drt^\tau$\REC, there is the need to introduce the auxiliary unary 
\emph{one-time-slice shift} operator $\shift$.
This operator can be explained as follows: $\shift(t)$, where $t$ is a
closed \ACP$_\drt^\tau$\REC\ term, denotes the process that remains 
after the process denoted by $t$ has idled till the next time slice.
The defining axioms for the one-time-slice  shift operator are given in 
Table~\ref{axioms-shift}.
\begin{table}[!t]
\caption{Defining axioms of the one-time-slice shift operator}
\label{axioms-shift}
\begin{eqntbl}
\begin{axcol}
\shift(\cts{a}) = \cts{\dead}                       & \axiom{DRSH1} \\
\shift(x \altc y) = \shift(x) \altc \shift(y)       & \axiom{DRSH2} \\
\shift(x \seqc y) = \shift(x) \seqc y               & \axiom{DRSH3} \\
\shift(\delay(x)) = x                               & \axiom{DRSH4} 
\end{axcol}
\end{eqntbl}
\end{table}%
In this table, $a$ stands for an arbitrary member of $\Acttd$.

We use the notation $\shift^n(t)$, where $n \in \Nat$, for the 
$n$-fold application of $\shift$ to $t$, i.e.\ $\shift^0(t) = t$ and 
$\shift^{n+1}(t) = \shift(\shift^{n}(t))$.

The relations from the time-stamped structural operational semantics of 
\ACP$_\drt^\tau$\REC\ are the smallest relations satisfying the rules 
given in Table~\ref{rules-ACPdrttaurec-ts}.
\begin{table}[p]
\caption{Transition rules for \ACP$_\drt^\tau$\REC}
\label{rules-ACPdrttaurec-ts}
\begin{ruletbl}
\Rule
{\phantom{\idling{x}{0}}}
{\idling{x}{0}}
\\
\Rule
{\phantom{\trans{x}{a}{x'}}}
{\trans{\cts{a}}{a[0]}{\surd}}
\\
\Rule
{\trans{x}{a[n]}{x'}}
{\trans{x \altc y}{a[n]}{x'}}
\quad
\Rule
{\trans{y}{a[n]}{y'}}
{\trans{x \altc y}{a[n]}{y'}}
\quad
\Rule
{\trans{x}{a[n]}{\surd}}
{\trans{x \altc y}{a[n]}{\surd}}
\quad
\Rule
{\trans{y}{a[n]}{\surd}}
{\trans{x \altc y}{a[n]}{\surd}}
\quad
\Rule
{\idling{x}{n}}
{\idling{x \altc y}{n}}
\quad
\Rule
{\idling{y}{n}}
{\idling{x \altc y}{n}}
\\
\Rule
{\trans{x}{a[n]}{x'}}
{\trans{x \seqc y}{a[n]}{x' \seqc y}}
\quad
\Rule
{\trans{x}{a[n]}{\surd}}
{\trans{x \seqc y}{a[n]}{y}}
\quad
\Rule
{\idling{x}{n}}
{\idling{x \seqc y}{n}}
\\
\Rule
{\trans{x}{a[n]}{x'}}
{\trans{\delay(x)}{a[n+1]}{x'}}
\quad
\Rule
{\trans{x}{a[n]}{\surd}}
{\trans{\delay(x)}{a[n+1]}{\surd}}
\quad
\Rule
{\idling{x}{n}}
{\idling{\delay(x)}{n+1}}
\\
\Rule
{\trans{x}{a[n]}{x'},\; \idling{y}{n}}
{\trans{x \parc y}{a[n]}{x' \parc \shift^n(y)}}
\quad
\Rule
{\idling{x}{n},\; \trans{y}{a[n]}{y'}}
{\trans{x \parc y}{a[n]}{\shift^n(x) \parc y'}}
\quad
\Rule
{\trans{x}{a[n]}{\surd},\; \idling{y}{n}}
{\trans{x \parc y}{a[n]}{\shift^n(y)}}
\quad
\Rule
{\idling{x}{n},\; \trans{y}{a[n]}{\surd}}
{\trans{x \parc y}{a[n]}{\shift^n(x)}}
\\
\RuleC
{\trans{x}{a[n]}{x'},\; \trans{y}{b[n]}{y'}}
{\trans{x \parc y}{c[n]}{x' \parc y'}}
{\commf(a,b) = c}
\quad
\RuleC
{\trans{x}{a[n]}{x'},\; \trans{y}{b[n]}{\surd}}
{\trans{x \parc y}{c[n]}{x'}}
{\commf(a,b) = c}
\\
\RuleC
{\trans{x}{a[n]}{\surd},\; \trans{y}{b[n]}{y'}}
{\trans{x \parc y}{c[n]}{y'}}
{\commf(a,b) = c}
\quad
\RuleC
{\trans{x}{a[n]}{\surd},\; \trans{y}{b[n]}{\surd}}
{\trans{x \parc y}{c[n]}{\surd}}
{\commf(a,b) = c}
\quad
\Rule
{\idling{x}{n},\; \idling{y}{n}}
{\idling{x \parc y}{n}}
\\
\Rule
{\trans{x}{a[n]}{x'},\; \idling{y}{n}}
{\trans{x \leftm y}{a[n]}{x' \parc \shift^n(y)}}
\quad
\Rule
{\trans{x}{a[n]}{\surd},\; \idling{y}{n}}
{\trans{x \leftm y}{a[n]}{\shift^n(y)}}
\quad
\Rule
{\idling{x}{n},\; \idling{y}{n}}
{\idling{x \leftm y}{n}}
\\
\RuleC
{\trans{x}{a[n]}{x'},\; \trans{y}{b[n]}{y'}}
{\trans{x \commm y}{c[n]}{x' \parc y'}}
{\commf(a,b) = c}
\quad
\RuleC
{\trans{x}{a[n]}{x'},\; \trans{y}{b[n]}{\surd}}
{\trans{x \commm y}{c[n]}{x'}}
{\commf(a,b) = c}
\\
\RuleC
{\trans{x}{a[n]}{\surd},\; \trans{y}{b[n]}{y'}}
{\trans{x \commm y}{c[n]}{y'}}
{\commf(a,b) = c}
\quad
\RuleC
{\trans{x}{a[n]}{\surd},\; \trans{y}{b[n]}{\surd}}
{\trans{x \commm y}{c[n]}{\surd}}
{\commf(a,b) = c}
\quad
\Rule
{\idling{x}{n},\; \idling{y}{n}}
{\idling{x \commm y}{n}}
\\
\RuleC
{\trans{x}{a[n]}{x'}}
{\trans{\encap{H}(x)}{a[n]}{\encap{H}(x')}}
{a \not\in H}
\quad
\RuleC
{\trans{x}{a[n]}{\surd}}
{\trans{\encap{H}(x)}{a[n]}{\surd}}
{a \not\in H}
\quad
\Rule
{\idling{x}{n}}
{\idling{\encap{H}(x)}{n}}
\\
\RuleC
{\trans{x}{a[n]}{x'}}
{\trans{\abstr{I}(x)}{a[n]}{\abstr{I}(x')}}
{a \not\in I}
\quad
\RuleC
{\trans{x}{a[n]}{\surd}}
{\trans{\abstr{I}(x)}{a[n]}{\surd}}
{a \not\in I}
\\
\RuleC
{\trans{x}{a[n]}{x'}}
{\trans{\abstr{I}(x)}{\tau[n]}{\abstr{I}(x')}}
{a \in I}
\quad
\RuleC
{\trans{x}{a[n]}{\surd}}
{\trans{\abstr{I}(x)}{\tau[n]}{\surd}}
{a \in I}
\quad
\Rule
{\idling{x}{n}}
{\idling{\abstr{I}(x)}{n}}
\\
\Rule
{\trans{x}{a[0]}{x'}}
{\trans{\timeout(x)}{a[0]}{x'}}
\quad
\Rule
{\trans{x}{a[0]}{\surd}}
{\trans{\timeout(x)}{a[0]}{\surd}}
\\
\RuleC
{\trans{\rec{t}{E}}{a[n]}{x'}}
{\trans{\rec{X}{E}}{a[n]}{x'}}
{X \!\!=\! t \,\in\, E}
\quad
\RuleC
{\trans{\rec{t}{E}}{a[n]}{\surd}}
{\trans{\rec{X}{E}}{a[n]}{\surd}}
{X \!\!=\! t \,\in\, E}
\quad
\RuleC
{\idling{\rec{t}{E}}{n}}
{\idling{\rec{X}{E}}{n}}
{X \!\!=\! t \,\in\, E}
\\
\Rule
{\trans{x}{a[n+1]}{x'}}
{\trans{\shift(x)}{a[n]}{x'}}
\quad
\Rule
{\trans{x}{a[n+1]}{\surd}}
{\trans{\shift(x)}{a[n]}{\surd}}
\quad
\Rule
{\idling{x}{n+1}}
{\idling{\shift(x)}{n}}
\end{ruletbl}
\end{table}%
In this table, 
$a$, $b$, and $c$ stand for arbitrary actions from $\Actt$,
$n$ stands for an arbitrary natural number from $\Nat$,
$H$ and $I$ stand for arbitrary subsets of $\Act$,
$X$ stands for an arbitrary variable from $\cX$, 
$t$ stands for an arbitrary \ACP$_\drt^\tau$\REC\ term, and
$E$ stands for an arbitrary guarded recursive specification over 
\ACP$_\drt^\tau$\REC.

In Figure~\ref{fig-time-stamped}, the processes that are denoted by the
closed terms
$\delay(\cts{a}) \altc \delay(\cts{b})$ and 
$\cts{\tau} \seqc \delay(\cts{a}) \altc \delay(\cts{b})$ 
according to the time-stamped operational semantics are presented 
graphically as far as the transitions are concerned.
Noteworthy is that, different from the processes denoted by these
terms according to the time-stamped operational semantics 
(see Figure~\ref{fig-two-phase}), both processes now have a choice.
\begin{figure}[!h]
\centering
\begin{tsfig}
\begin{array}[t]{c}
\delay(\cts{a}) \altc \delay(\cts{b}) \\
\stepsw{a[1]} \stepse{b[1]} \\
\surd \phantom{\stepsw{} \quad \stepse{}} \surd
\end{array}
\qquad \qquad
\begin{array}[t]{c}
\cts{\tau} \seqc \delay(\cts{a}) \altc \delay(\cts{b}) \\
\stepsw{\tau[0]} \stepse{b[1]} \\
\begin{array}[t]{c@{}}
\delay(\cts{a})  \\
\stepdl{a[1]} \\
\surd
\end{array}
\phantom{\stepsw{} \;\;\; \stepse{}}
\begin{array}[t]{@{}c}
\surd
\end{array}
\end{array}
\end{tsfig}
\caption{Graphical time-stamped presentation of two processes}
\label{fig-time-stamped}
\end{figure}

A \emph{ts-branching bisimulation} is a binary relation 
$R$ on $\PTerm \union \set{\surd}$ such that, 
for all $t_1,t_2 \in \PTerm$ with $R(t_1,t_2)$, 
the following transfer conditions hold:
\begin{itemize}
\item 
if $\trans{t_1}{a[n]}{t_1'}$, then
\begin{itemize}
\item 
either 
$a \equiv \tau$ and $R(t_1',\shift^n(t_2))$ 
\item 
or, for some $m \in \Nat$, there exist $t^*_0,\ldots,t^*_m \in \PTerm$, 
$t_2' \in \PTerm \union \set{\surd}$, and $n_0,\ldots,n_m \in \Nat$ 
with $t^*_0 \equiv t_2$ and $\sum_{i=0}^m n_i = n$ such that
\begin{itemize}
\item 
for all $k \in \Nat$ with $k < m$, \\ \hspace*{1em}
$\trans{t^*_k}{\tau[n_k]}{t^*_{k+1}}$ and 
$R(\shift^{\sum_{i=0}^k n_i}(t_1),t^*_{k+1})$;
\item 
$\trans{t^*_m}{a[n_m]}{t_2'}$, and $R(t_1',t_2')$;
\end{itemize}
\end{itemize}
\item 
if $\trans{t_2}{a[n]}{t_2'}$, then
\begin{itemize}
\item 
either 
$a \equiv \tau$ and if $R(\shift^n(t_1),t_2')$ 
\item 
or, for some $m \in \Nat$, there exist $t^*_0,\ldots,t^*_m \in \PTerm$,
$t_1' \in \PTerm \union \set{\surd}$, and $n_0,\ldots,n_m \in \Nat$
with $t^*_0 \equiv t_1$ and $\sum_{i=0}^m n_i = n$ such that 
\begin{itemize}
\item 
for all $k \in \Nat$ with $k < m$, \\ \hspace*{1em}
$\trans{t^*_k}{\tau[n_k]}{t^*_{k+1}}$ and 
$R(t^*_{k+1},\shift^{\sum_{i=0}^k n_i}(t_2))$;
\item
$\trans{t^*_m}{a[n_m]}{t_1'}$, and $R(t_1',t_2')$;
\end{itemize}
\end{itemize}
\item
$\idling{t_1}{n}$ iff $\idling{t_2}{n}$.
\end{itemize}
Two terms $t_1,t_2 \in \PTerm$ are \emph{ts-branching bisimilar}, 
written $t_1 \bisim_\br' t_2$, if there exists a ts-branching 
bisimulation $R$ with $R(t_1,t_2)$.

If $R$ is a binary relation on $\PTerm \union \set{\surd}$, then a pair 
$(t_1,t_2) \in \PTerm \x \PTerm$ is said to satisfy the 
\emph{ts-root condition} in $R$ if the following condition 
holds:
\begin{itemize}
\item 
if $\trans{t_1}{a[n]}{t_1'}$, then 
there exists a $t_2' \in \PTerm \union \set{\surd}$ such that 
\smash{$\trans{t_2}{a[n]}{t_2'}$} and $R(t_1',t_2')$;
\item 
if $\trans{t_2}{a[n]}{t_2'}$, then 
there exists a $t_1' \in \PTerm \union \set{\surd}$ such that 
\smash{$\trans{t_1}{a[n]}{t_1'}$} and $R(t_1',t_2')$.
\end{itemize}
Two terms $t_1,t_2 \in \PTerm$ are \emph{rooted ts-branching bisimilar}, 
written $t_1 \bisim_\rb' t_2$, if there exists a ts-branching 
bisimulation $R$ such that $R(t_1,t_2)$ and $(t_1,t_2)$ satisfies the 
ts-root condition in $R$.

Rooted ts-branching bisimilarity is an equivalence relation.
\begin{theorem}[Equivalence]
\label{theorem-ts-rbbisim-equiv}
The binary relation $\bisim_\rb'$ on $\PTerm$ is an equivalence relation.
\end{theorem}
\begin{proof}
The proofs of reflexivity and symmetry of $\bisim_\rb'$ go the same as 
the proofs of reflexivity and symmetry of $\bisim_\rb$.
To prove that $t \bisim_\rb' t'$ and $t' \bisim_\rb' t''$ implies  
$t \bisim_\rb' t''$, consider the relation composition of witnessing 
branching bisimulations $R$ and $R'$ for $t \bisim_\rb' t'$ and 
$t' \bisim_\rb' t''$, respectively.
Take arbitrary $t_1,t_2,t_3 \in \PTerm$ such that $R(t_1,t_2)$ and
$R'(t_2,t_3)$.
Two cases have to be distinguished:
(a)~the case where $\tup{t_1,t_2}$ has to satisfy the root condition in 
$R$ and
(b)~the case where $\tup{t_1,t_2}$ does not have to satisfy the root 
condition in $R$. 
If $\tup{t_1,t_2}$ has to satisfy the root condition in $R$, then 
$\tup{t_2,t_3}$ has to satisfy the root condition in $R'$ as well.
Case~(a) follows immediately from the satisfaction of the the root 
condition in $R$ and $R'$ by $\tup{t_1,t_2}$ and $\tup{t_2,t_3}$, 
respectively.
The proof of case~(b) goes along the same lines as the proof of 
Proposition~2.11 from~\cite{GW96a} given in~\cite{Bas96a}, but taking 
into account the number of time slices involved in sequences of silent 
steps.
Such a similar proof is possible because the replacement of the 
conditions
\begin{itemize}
\item
for all $k \in \Nat$ with $k < m$, \\ \hspace*{1em} 
$\trans{t^*_k}{\tau[n_k]}{t^*_{k+1}}$ and 
$R(\shift^{\sum_{i=0}^k n_i}(t_1),t^*_{k+1})$;
\item
for all $k \in \Nat$ with $k < m$, \\ \hspace*{1em} 
$\trans{t^*_k}{\tau[n_k]}{t^*_{k+1}}$ and 
$R(t^*_{k+1},\shift^{\sum_{i=0}^k n_i}(t_2))$
\end{itemize}
in the definition of ts-branching bisimulation by
\begin{itemize}
\item
for all $k \in \Nat$ with $k < m$, \\ \hspace*{1em} 
$\trans{t^*_k}{\tau[n_k]}{t^*_{k+1}}$ and 
if $k + 1 = m$ then
$R(\shift^{\sum_{i=0}^k n_i}(t_1),t^*_{k+1})$;
\item
for all $k \in \Nat$ with $k < m$, \\ \hspace*{1em} 
$\trans{t^*_k}{\tau[n_k]}{t^*_{k+1}}$ and 
if $k + 1 = m$ then
$R(t^*_{k+1},\shift^{\sum_{i=0}^k n_i}(t_2))$
\end{itemize}
yields a bisimilarity relation that coincides with ts-branching 
bisimilarity as defined above.
This is due to a stuttering property that can be proved in the same way 
as Lemma~2.5 from~\cite{GW96a}, but taking into account the number of 
time slices involved in sequences of silent steps.
\qed
\end{proof}
The fact that rooted ts-branching bisimilarity is an equivalence 
relation also follows from Theorem~\ref{theorem-eqv-def} (see below).
The proof of Theorem~\ref{theorem-ts-rbbisim-equiv} given above shows 
that this fact does not depend on the extension of the transition 
relations and idling relations involved.

In Figure~\ref{fig-two-phase-time-stamped}, the process that is denoted 
by the closed term $\delay(\delay(\cts{a})) \altc \delay(\cts{b})$ is 
presented graphically as far as the transitions are concerned, both 
according to the two-phase operational semantics and according to the 
time-stamped operational semantics.
In this example, the two presentations convey in some sense the same 
behaviour.
This raises the question how the two-phase structural operational 
semantics of \ACP$_\drt^\tau$\REC\ and the time-stamped structural 
operational semantics of \ACP$_\drt^\tau$\REC\ are exactly related.
\begin{figure}[!h]
\centering
\begin{tsfig}
\begin{array}[t]{c}
\delay(\delay(\cts{a})) \altc \delay(\cts{b}) \\
\stepdl{\sigma} \\
\delay(\cts{a}) \altc \cts{b} \\
\stepsw{\sigma} \stepse{b} \\
\begin{array}[t]{c@{}}
\cts{a} \\
\stepdl{a} \\
\surd
\end{array}
\phantom{\stepsw{} \;\; \stepse{}}
\begin{array}[t]{@{}c}
\surd
\end{array}
\end{array}
\qquad \qquad
\begin{array}[t]{c}
\delay(\delay(\cts{a})) \altc \delay(\cts{b}) \\
\stepsw{a[2]} \stepse{b[1]} \\
\begin{array}[t]{c@{}}
\surd
\end{array}
\phantom{\stepsw{} \quad\;\; \stepse{}}
\begin{array}[t]{@{}c}
\surd
\end{array}
\end{array}
\end{tsfig}
\caption{Graphical two-phase and time-stamped presentation of one process}
\label{fig-two-phase-time-stamped}
\end{figure}

Below, we write \smash{$\trans{t}{\sigma^{n+1}}{t'}$}, where 
$t, t' \in \PTerm$ and $n \in \Nat$, to indicate that 
there exist $t_0,\ldots,t_{n+1} \in \PTerm$ with $t_0 \equiv t$ and
$t_{n+1} \equiv t'$ such that, for all $i \in \Nat$ with $i \leq n$, 
$\trans{t_i}{\sigma}{t_{i+1}}$. 

The following proposition relates the two-phase structural operational 
semantics of \ACP$_\drt^\tau$\REC\ and the time-stamped structural 
operational semantics of \ACP$_\drt^\tau$\REC.
\begin{proposition}
\label{proposition-rel-tp-ts}
For all $t \in \PTerm$, $t' \in \PTerm \union \set{\surd}$, 
$a \in \Actt$, and $n \in \Nat$, the following holds:
\begin{itemize}
\item[(1)]
$\trans{t}{a[0]}{t'}$ iff $\trans{t}{a}{t'}$;
\item[(2)]
$\trans{t}{a[n+1]}{t'}$ iff 
there exists a $t'' \in \PTerm$ such that 
$\trans{t}{\sigma^{n+1}}{t''}$ and $\trans{t''}{a}{t'}$;
\item[(3)]
$\idling{t}{n+1}$ iff 
there exists a $t'' \in \PTerm$ such that 
$\trans{t}{\sigma^{n+1}}{t''}$.
\end{itemize}
\end{proposition}
\begin{proof}
Both the `if' part and the `only if' part of (1)--(3) are easily proved 
by induction on the structure of $t$.
\qed
\end{proof}

Rooted branching bisimilarity and rooted ts-branching bisimilarity 
coincide.%
\begin{theorem}[Equivalent definitions]
\label{theorem-eqv-def}
For all terms $t_1, t_2 \in \PTerm$, $t_1 \bisim_\rb t_2$ iff 
$t_1 \bisim_\rb' t_2$.
\end{theorem}
\begin{proof}
The `only if' part.
Assume that $t_1, t_2 \in \PTerm$ and $t_1 \bisim_\rb t_2$.
Then there is a branching bisimulation $R$ witnessing 
$t_1 \bisim_\rb t_2$.
Construct a binary relation $R'$ on $\PTerm \union \set{\surd}$ from $R$
as follows:
\begin{ldispl}
R' = R \diff 
\Union_{n \in \Nat} 
 \set{\tup{t_1'',t_2''} \where 
      \Exists{t_1',t_2' \in \PTerm}
       {R(t_1',t_2') \Land
        \trans{t_1'}{\sigma^{n+1}}{t_1''} \Land
        \trans{t_2'}{\sigma^{n+1}}{t_2''}}}\;.
\end{ldispl}%
Using Proposition~\ref{proposition-rel-tp-ts}, it is straightforward to 
check that the constructed relation $R'$ is a ts-branching bisimulation 
witnessing $t_1 \bisim_\rb' t_2$.

The `if' part.
Assume that $t_1, t_2 \in \PTerm$ and $t_1 \bisim_\rb' t_2$.
Then there is a ts-branching bisimulation $S$ witnessing 
$t_1 \bisim_\rb' t_2$.
Construct a binary relation $S'$ on $\PTerm \union \set{\surd}$ from $S$
as follows:
\begin{ldispl}
S' = S \union 
\Union_{n \in \Nat} 
 \set{\tup{t_1'',t_2''} \where 
      \Exists{t_1',t_2' \in \PTerm}
       {S(t_1',t_2') \Land
        \trans{t_1'}{\sigma^{n+1}}{t_1''} \Land
        \trans{t_2'}{\sigma^{n+1}}{t_2''}}}\;.
\end{ldispl}%
Using Proposition~\ref{proposition-rel-tp-ts}, it is  straightforward to 
check that the constructed relation $S'$ is a branching bisimulation 
witnessing $t_1 \bisim_\rb t_2$.
\qed
\end{proof}
It may come across as surprising that not too much is removed from $R$ 
in the proof of the `only if' part of Theorem~\ref{theorem-eqv-def}.
However, that not too much is removed is self-evident when realising 
that the following is a direct corollary of 
Proposition~\ref{proposition-rel-tp-ts}:
there exist $s'',t'' \in \PTerm$ such that 
\smash{$\trans{t}{\sigma^{m+1}}{s''}$}, 
\smash{$\trans{s''}{\sigma^{n-m}}{t''}$}, 
$\trans{t''}{a}{t'}$, \linebreak[2] and $\trans{s''}{b}{s'}$ iff
$\trans{t}{a[n+1]}{t'}$, $\trans{t}{b[m+1]}{s'}$, and 
$m < n$.
Figure~\ref{fig-two-phase-time-stamped} above gives an example where 
$n = 2$ and $m = 1$.

The following soundness result is a corollary of 
Theorems~\ref{theorem-soundness} and~\ref{theorem-eqv-def}.
\begin{corollary}
\label{corollary-soundness}
For all terms $t, t' \in \PTerm$, $t = t'$ is derivable from the axioms 
of \ACP$_\drt^\tau$\REC\ only if $t \bisim_\rb' t'$.
\end{corollary}

Basic terms as defined in Section~\ref{subsect-acpdrt} reflect the 
two-phase semantics of \ACP$_\drt^\tau$.
A variant of basic terms, called ts-basic terms, that reflect the 
time-stamped semantics of \ACP$_\drt^\tau$ can also be defined.
The set $\cB'$ of \emph{ts-basic terms} over \ACP$_\drt^\tau$ is the 
smallest set satisfying:
\begin{itemize}
\item
if $a \in \Acttd$ and $n \in \Nat$, 
then $\sigma^n(\cts{a}) \in \cB'$;
\item
if $a \in \Actt$, $n \in \Nat$, and $t \in \cB'$, 
then $\sigma^n(\cts{a}) \seqc t \in \cB'$;
\item
if $t,t' \in \cB'$, then $t \altc t' \in \cB'$.
\end{itemize}
Modulo axioms A1 and A2, each ts-basic term from $\cB'$ is of the 
following form:
\begin{ldispl}
\Altc{1 \leq i \leq n} \delay^{n_i}(\cts{a_i}) \seqc t_i \altc
\Altc{1 \leq j \leq m} \delay^{m_j}(\cts{b_j})\;,
\end{ldispl}%
where
$n,m \in \Nat$,
$a_i \in \Actt$, $n_i \in \Nat$, and $t_i \in \cB'$ for 
$1 \leq i \leq n$, and
$b_j \in \Actt$ and $m_i \in \Nat$ for $1 \leq j \leq m$. 

The following proposition states that each closed \ACP$_\drt^\tau$ term 
can be reduced to a ts-basic term.
\begin{proposition}
\label{proposition-elim}
For each closed term $t$ of \ACP$_\drt^\tau$, there exists a ts-basic 
term $t' \in \cB'$ such that $t = t'$ is derivable from the axioms of 
\ACP$_\drt^\tau$.
\end{proposition}
\begin{proof}
By Theorem~\ref{theorem-elim}, it is sufficient to prove the following:
\begin{quote}
\it
for each basic term $t \in \cB$, there exists a ts-basic term 
$t' \in \cB'$ such that $t = t'$ is derivable from the axioms of 
\ACP$_\drt^\tau$.
\end{quote}
This is easily proved by induction on the structure of $t$.
\qed
\end{proof}

\section{A Coarser Version of Branching Bisimilarity}
\label{sect-coarser-eqv}

The axioms of \ACP$_\drt^\tau$ concerning silent steps (DRB1--DRB4) are 
based on the standard notion of branching bisimilarity for processes 
with discrete relative timing, i.e.\ the one first introduced 
in~\cite{BB95b}.
The simple idea behind this notion of branching bisimilarity is that an 
internal action can be forgotten wherever no options are lost by 
performing it.
The experience with analyzing the PAR protocol (and other protocols) 
triggered a quest for a version of branching bisimilarity for processes 
with discrete relative timing that is coarser than the standard version.
In this section, a variant of the standard version will be defined that 
is coarser than the standard version by treating an internal action
always as redundant if it is followed by a process that is only capable 
of idling till the next time slice.

In~\cite{BMR02b}, a first attempt has been made to define the same 
variant.
However, the attempt was unsatisfactory.
The definition of the variant in that paper was in fact complicated to 
such an extent that it led to errors in the paper that were not spotted 
at the time.
Like the definition of the standard version, the definition con\-cerned 
refers to a two-phase operational semantics of a version of \ACP\ with 
abstraction for processes with discrete relative timing.
In this section, a definition of the variant will be given that refers 
to the time-stamped operational semantics of \ACP$_\drt^\tau$ presented 
in Section~\ref{subsect-standard-time-stamped}.
It turns out that taking a time-stamped operational semantics as the 
basis results in a definition that is far less complicated than the one 
from~\cite{BMR02b}.

\subsection{Dormancy-Aware Branching Bisimilarity}
\label{subsect-coarser-eqv}

In this section, a version of branching bisimilarity for processes with 
discrete relative timing is introduced that is coarser than the one 
introduced in Section~\ref{sect-standard-bisim}.
The version introduced is coarser as a result of treating an internal 
action always as redundant if it is followed by a process that is only 
capable of idling till the next time slice.

With regard to the motivation for this coarser version, consider 
(a)~a process $p$ with the option of idling till the next time slice and 
then to behave as a process~$q$ and
(b)~a process $p'$ that behaves as $p$ except that in the 
above-mentioned option the idling till the next time slice is preceded 
by performing an internal action in the current time slice.
The capabilities of $p$ change by idling till the next time slice and 
the capabilities of $p'$ change by performing an internal action in the 
current time slice. 
However, since it requires unlikely means to perceive differences 
between spending time in performing an action that is considered to be 
unobservable and spending time in not performing any action, the latter 
change is considered to be observable only after the subsequent idling 
till the next time slice.
In either case, the capabilities after the idling till the next time 
slice are the same.
In other words, it is hard to imagine a realistic experiment that can 
distinguish between the processes $p$ and $p'$.

To enforce that the passage of time cannot introduce non-determinism, 
the two-phase structural operational semantics of \ACP$_\drt^\tau$\REC\ 
is such that, for all $t,t' \in \PTerm$, $\delay(t) \altc \delay(t')$ 
can only make the transition 
$\trans{\delay(t) \altc \delay(t')}{\sigma}{t \altc t'}$.
This way to enforce that the passage of time cannot introduce 
non-determinism causes a problem with the treatment of an internal 
action as redundant if it is followed by a process that is only capable 
of idling till the next time slice:
even if a version of branching bisimilarity allows of forgetting about 
the internal action involved, $\tau \seqc \delay(t) \altc \delay(t')$ 
can still make two transitions (see also Figure~\ref{fig-two-phase}). 
This problematic difference in the number of possible transitions does 
not occur with the time-stamped structural operational semantics of 
\ACP$_\drt^\tau$\REC\ (see also Figure~\ref{fig-time-stamped}).
Therefore, the definition of the coarser version of branching 
bisimilarity will refer to the time-stamped structural operational 
semantics of \ACP$_\drt^\tau$\REC\ instead.

If a process is capable of performing an action in the current time 
slice, then it is regarded as being active.
We inductively define a unary \emph{activeness} relation $\cA$ on 
$\PTerm \union \set{\surd}$ as follows:
\begin{itemize}
\item 
if $\trans{t}{a[0]}{t'}$\, for some $a \in \Act$ and $t' \in \PTerm$, 
then $\cA(t)$;
\item 
if $\trans{t}{a[0]}{\surd}$ for some $a \in \Actt$, then $\cA(t)$;
\item
if $\trans{t}{\tau[0]}{t'}$\, and $\cA(t')$, then $\cA(t)$.
\end{itemize}
If not $\cA(t)$, then the process denoted by $t$ is said to be dormant.

A \emph{dormancy-aware} branching bisimulation is a binary relation 
$R$ on {$\PTerm \union\nolinebreak \set{\surd}$} such that, 
for all $t_1,t_2 \in \PTerm$ with $R(t_1,t_2)$, 
the following transfer conditions hold:
\begin{itemize}
\item 
if $\trans{t_1}{a[n]}{t_1'}$, then
\begin{itemize}
\item 
either 
$a \equiv \tau$ and if $\cA(t_1')$ then $R(t_1',\shift^n(t_2))$ 
\item 
or, for some $m \in \Nat$, there exist $t^*_0,\ldots,t^*_m \in \PTerm$, 
$t_2' \in \PTerm \union \set{\surd}$, and $n_0,\ldots,n_m \in \Nat$ 
with $t^*_0 \equiv t_2$ and $\sum_{i=0}^m n_i = n$ such that
\begin{itemize}
\item 
for all $k \in \Nat$ with $k < m$, \\ \hspace*{1em}
$\trans{t^*_k}{\tau[n_k]}{t^*_{k+1}}$ and if $\cA(t^*_{k+1})$ then 
$R(\shift^{\sum_{i=0}^k n_i}(t_1),t^*_{k+1})$;
\item 
$\trans{t^*_m}{a[n_m]}{t_2'}$, and $R(t_1',t_2')$;
\end{itemize}
\end{itemize}
\item 
if $\trans{t_2}{a[n]}{t_2'}$, then
\begin{itemize}
\item 
either 
$a \equiv \tau$ and if $\cA(t_2')$ then $R(\shift^n(t_1),t_2')$ 
\item 
or, for some $m \in \Nat$, there exist $t^*_0,\ldots,t^*_m \in \PTerm$,
$t_1' \in \PTerm \union \set{\surd}$, and $n_0,\ldots,n_m \in \Nat$
with $t^*_0 \equiv t_1$ and $\sum_{i=0}^m n_i = n$ such that 
\begin{itemize}
\item 
for all $k \in \Nat$ with $k < m$, \\ \hspace*{1em}
$\trans{t^*_k}{\tau[n_k]}{t^*_{k+1}}$ and if $\cA(t^*_{k+1})$ then 
$R(t^*_{k+1},\shift^{\sum_{i=0}^k n_i}(t_2))$;
\item
$\trans{t^*_m}{a[n_m]}{t_1'}$, and $R(t_1',t_2')$;
\end{itemize}
\end{itemize}
\item
$\idling{t_1}{n}$ iff $\idling{t_2}{n}$.
\end{itemize}
The two processes presented graphically in Figure~\ref{fig-time-stamped}
are a simple example of processes that are dormancy-aware branching 
bisimilar but not ts-branching bisimilar.

Two terms $t_1,t_2 \in \PTerm$ are rooted \emph{dormancy-aware} 
branching bisimilar, written $t_1 \dabisim_\rb t_2$, if there exists a 
dormancy-aware branching bisimulation $R$ such that $R(t_1,t_2)$ 
and $(t_1,t_2)$ satisfies the ts-root condition in $R$.

Adoption of rooted dormancy-aware branching bisimilarity leads to 
the addition of the axiom given in Table~\ref{axiom-DRB5} to the axiom 
system of \BPA$^\drt_\tau$\REC.
\begin{table}[!t]
\caption{Additional axiom for dormancy-aware branching bisimilarity}
\label{axiom-DRB5}
\begin{eqntbl}
\begin{axcol}
\cts{a} \seqc (\delay^n(\cts{\tau} \seqc \delay(x)) \altc y) = 
\cts{a} \seqc (\delay^n(\delay(x)) \altc y)               & \axiom{DRB5} 
\end{axcol}
\end{eqntbl}
\end{table}
In this table, 
$a$ stands for an arbitrary actions from $\Actt$ and
$n$ stands for an arbitrary natural number from $\Nat$.

Axiom DRB5 is the axiom that characterizes the additional 
identifications made by rooted dormancy-aware branching 
bisimilarity.
In Section~\ref{subsect-properties}, we will come back to the point that
axiom DRB5 is valid with respect to rooted dormancy-aware branching 
bisimilarity.

\subsection{Properties of Rooted Dormancy-Aware Branching Bisimilarity}
\label{subsect-properties}

Rooted dormancy-aware branching bisimilarity is coarser than rooted 
branching bisimilarity.
\begin{theorem}[Inclusion]
\label{theorem-coarser}
Let $t$ and $t'$ be closed terms of \ACP$_\drt^\tau$\REC.
Then $t \bisim_\rb t'$ only if $t \dabisim_\rb t'$.
\end{theorem}
\begin{proof}
This follows directly from the definitions of $\bisim_\rb$ and 
$\dabisim_\rb$.
\qed
\end{proof}

Rooted dormancy-aware branching bisimilarity is an equivalence relation.
\begin{theorem}[Equivalence]
The relation $\dabisim_\rb$ is an equivalence relation.
\end{theorem}
\begin{proof}
Define \emph{generalized silent step} relations 
$\gsstep{\ph}{\tau[n]}{\ph} \subseteq
 \PTerm \x (\PTerm \union \set{\surd})$
for $n \in \Nat$ as the smallest relations satisfying:
\begin{itemize}
\item
if $\trans{t}{\tau[n]}{t'}$ and $\cA(t')$, 
then $\gsstep{t}{\tau[n]}{t'}$;
\item
if $\trans{t}{\tau[n]}{t'}$, $\gsstep{t'}{\tau[n']}{t''}$, and
not $\cA(t')$, then $\gsstep{t}{\tau[n+n']}{t''}$.
\end{itemize} 
Reformulate the definition of rooted dormancy-aware branching 
bisimilarity that refers to the time-stamped operational semantics 
defined in Section~\ref{subsect-standard-time-stamped} using the 
generalized silent step relations.
Then the reformulated definition is also a definition of rooted 
ts-branching bisimilarity that refers to the adapted time-stamped 
operational semantics of \ACP$_\drt^\tau$\REC\ that consists of:
\begin{itemize}
\item
the binary transition relation
$\trans{\ph}{a[n]}{\ph} \subseteq \PTerm \x (\PTerm \union \set{\surd})$
defined in Section~\ref{subsect-standard-time-stamped} for each 
$a \in \Act$ and $n \in \Nat$;
\item
the binary generalized silent step relation
$\gsstep{\ph}{\tau[n]}
        {\ph} \subseteq \PTerm \x (\PTerm \union \set{\surd})$
defined above for each $n \in \Nat$;
\item
the unary idling relation ${\idling{\ph}{n}} \subseteq \PTerm$ defined 
in Section~\ref{subsect-standard-time-stamped} for each $n \in \Nat$.
\end{itemize}
The proof of Theorem~\ref{theorem-ts-rbbisim-equiv} given in 
Section~\ref{subsect-standard-time-stamped} shows that the fact that 
rooted ts-branching bisimilarity is an equivalence relation does not 
depend on the extension of the transition relations and idling relations 
involved.
From this and the reformulated definition, it follows immediately that 
rooted dormancy-aware branching bisimilarity is an equivalence 
relation.
\qed
\end{proof}

Rooted dormancy-aware branching bisimilarity is not a congruence 
with respect to the operators $\parc$, $\leftm$, and $\commm$. 
For example, 
$\cts{a} \seqc \delay(\cts{b}) \dabisim_\rb 
 \cts{a} \seqc \cts{\tau} \seqc \delay(\cts{b})$ and
$\cts{c} \dabisim_\rb \cts{c}$, but 
$\cts{a} \seqc \delay(\cts{b}) \parc \cts{c} \not\dabisim_\rb 
 \cts{a} \seqc \cts{\tau} \seqc \delay(\cts{b}) \parc \cts{c}$
($\cts{c}$ may be performed after $\cts{\tau}$).
However, rooted dormancy-aware branching bisimilarity is preserved 
by all operators of \BPA$_\drt^\tau$\REC.
\begin{theorem}[Congruence]
The relation $\dabisim_\rb$ is a congruence relation with respect to all 
operators from the signature of \BPA$_\drt^\tau$\REC.
\end{theorem}
\begin{proof}
Let $t_1,t_1',t_2,t_2' \in \PTerm$ be such that $t_1 \dabisim_\rb t_1'$ 
and $t_2 \dabisim_\rb t_2'$, and 
let $R_1$ and $R_2$ be dormancy-aware branching bisimulations 
witnessing $t_1 \dabisim_\rb t_1'$ and $t_2 \dabisim_\rb t_2'$, 
respectively.
Define, for each operator $\diamond$ of \BPA$_\drt^\tau$\REC, a 
binary relation $R_\diamond$ on $\PTerm \union \set{\surd}$ as follows:
\begin{ldispl}
\begin{aeqns}
R_\altc    & = & 
\set{(t_1 \altc t_2,t_1' \altc t_2')} \union R_1 \union R_2\;,
\\
R_\seqc    & = & 
\set{(s_1 \seqc t_2,s_1' \seqc t_2') \where R_1(s_1,s_1')} \union R_2\;, 
\\
R_\delay   & = & 
\set{(\delay(t_1),\delay(t_1'))} \union R_1\;,
\\
R_\timeout & = & 
\set{(\timeout(t_1),\timeout(t_1'))} \union R_1\;.
\end{aeqns}
\end{ldispl}%
It is a routine matter to check that these relations satisfy the 
transfer and root conditions required to be dormancy-aware branching 
bisimulations witnessing
$t_1 \altc t_2 \dabisim_\rb t_1' \altc t_2'$,
$t_1 \seqc t_2 \dabisim_\rb t_1' \seqc t_2'$,
$\delay(t_1) \dabisim_\rb \delay(t_1')$, and
$\timeout(t_1) \dabisim_\rb \timeout(t_1')$, respectively.
\qed
\end{proof}

The axiom system of \BPA$_\drt^\tau$\REC\ extended with axiom DRB5 is 
sound with respect to $\dabisim_\rb$ for equations between terms from 
$\PTerm$ in which only constants and operators of \BPA$_\drt^\tau$\REC\ 
occur.
\begin{theorem}[Soundness]
\label{theorem-soundness-2}
For all terms $t, t' \in \PTerm$ that are closed terms of 
\BPA$_\drt^\tau$\REC, $t = t'$ is derivable from the axioms 
of \BPA$_\drt^\tau$\REC\ and/or axiom DRB5 only if $t \dabisim_\rb t'$.
\end{theorem}
\begin{proof}
Because $\dabisim_\rb$ is a congruence with respect to all operators from 
the signature of \BPA$_\drt^\tau$\REC, only the validity of each axiom 
of \BPA$_\drt^\tau$\REC\ has to be proved.
By Theorem~\ref{theorem-soundness}, the axioms of \BPA$_\drt^\tau$\REC\ 
are valid with respect to $\bisim_\rb$.
By Theorem~\ref{theorem-coarser}, validity with respect to $\bisim_\rb$ 
implies validity with respect to $\dabisim_\rb$.
Therefore, it is sufficient to prove the validity of the instances of
axiom schema DRB5 with respect to $\dabisim_\rb$.
The notation introduced in the proof of Theorem~\ref{theorem-soundness} 
is used here as well.

For each instance $\ax$ of axiom schema DRB5, a dormancy-aware 
branching bisimulation $R_\ax$ witnessing the validity of $\ax$ can be 
constructed as follows:
\begin{ldispl}
R_\ax = 
\set{\tup{t,t'} \where t = t' \in \csi(\ax)} 
\\ \phantom{R_\ax = {}}\,
 {} \union
 \set{\tup{t,t'} \where 
      t = t' \in 
      \csi(\delay^n(\cts{\tau} \seqc \delay(x)) \altc y = 
           \delay^n(\delay(x)) \altc y)} 
\\ \phantom{R_\ax = {}}\, 
 {}  \union 
 \set{\tup{t,t'} \where 
      t = t' \in \csi(\cts{\tau} \seqc \delay(x) = \delay(x))}
\union R_\id\;.
\end{ldispl}%
It is straightforward to check that the constructed relation $R_\ax$ is 
a dormancy-aware branching bisimulation witnessing, for each closed 
substitution instance $t = t'$ of $\ax$, $t \dabisim_\rb t'$.
\qed
\end{proof}
As in the case of the axiom system of \BPA$_\drt^\tau$\REC\ and 
$\bisim_\rb$, the axiom system of \BPA$_\drt^\tau$\REC\ extended with 
axiom DRB5 is not complete with respect to $\dabisim_\rb$ for equations 
between terms from $\PTerm$ in which only the constants and operators of 
\BPA$_\drt^\tau$\REC\ occur.
It is an open problem whether the axiom system of \BPA$_\drt^\tau$ 
extended with axiom DRB5 is complete with respect to $\dabisim_\rb$ for 
equations between terms from $\PTerm$ in which only the constants and 
operators of \BPA$_\drt^\tau$ occur.

\subsection{Verification in Two Phases}
\label{subsect-two-phase-verification}

If axiom DRB5 is possibly needed, a natural way to verify an equation of 
the form $\abstr{I}(t) = t'$, for $t,t' \in \PTerm$ where no abstraction 
operator occurs in $t$ and only constants and operators of 
\BPA$_\drt^\tau$\REC\ occur in $t'$, is the following two-phase way:
\begin{itemize}
\item
first, derive from the axioms of \ACP$_\drt^\tau$\REC\ an equation
$t = t''$ for some $t'' \in \PTerm$ in which only constants and 
operators of \BPA$_\drt^\tau$\REC\ occur;
\item
then, derive from the axioms of \BPA$_\drt^\tau$\REC\ and DRB5 the 
equation $\abstr{I}(t'') = t'$.
\end{itemize}
This raises the question whether the first phase is always possible.
This question can be answered in the affirmative.

\begin{theorem}
\label{theorem-verification}
For all terms $t \in \PTerm$ in which no abstraction operator occurs, 
there exists a term $t' \in \PTerm$ that is a closed term of 
\BPA$_\drt^\tau$\REC\ such that $t = t'$ is derivable from the axioms 
of \ACP$_\drt^\tau$\REC.
\end{theorem}
\begin{proof}
Define the set $\cL$ of \emph{linear terms} over \ACP$_\drt^\tau$ to be 
the smallest set satisfying:
\begin{itemize}
\item
if $a \in \Acttd$, then $\cts{a} \in \cL$;
\item
if $a \in \Actt$ and $X \in \cX$, then $\cts{a} \seqc X \in \cL$;
\item
if $X \in \cX$, then $\delay(X) \in \cL$;
\item
if $t,t' \in \cL$, then $t \altc t' \in \cL$.
\end{itemize}
Define a \emph{linear recursive specification} over 
\ACP$_\drt^\tau$\REC\ as a guarded recursive specification $E$ over
\ACP$_\drt^\tau$\REC\ where, for each equation $X \!= t\; \in \;E$,
$t \in \cL$.

Claim that, for each guarded recursive specification $E$ over 
\ACP$_\drt^\tau$\REC\ and each $X \in \vars(E)$, there exists a linear
recursive specification $E'$ over \ACP$_\drt^\tau$\REC\ such that
$\rec{X}{E} = \rec{X}{E'}$ is derivable from the axioms of 
\ACP$_\drt^\tau$\REC.
The proof of this claim goes essentially the same as the proof of 
Theorem~1 from~\cite{GM20a}.

Claim that, for all terms $t \in \PTerm$ in which no abstraction 
operator occurs, there exists a linear recursive specification $E$ over 
\ACP$_\drt^\tau$\REC\ and an $X \in \vars(E)$ such that $t = \rec{X}{E}$
is derivable from the axioms of \ACP$_\drt^\tau$\REC. 
The proof of this claim goes essentially the same as the proof of 
Lemma~1 from~\cite{Mid21a} except that the case where $t$ is a constant
$\rec{X}{E}$ now requires the use of the previous claim.
The proof involves constructions of linear recursive specifications from 
linear recursive specifications for the operators of 
\ACP$_\drt^\tau$\REC\ other than the abstraction operators. 
For the greater part, the constructions are reminiscent of operations on 
process graphs defined in Sections 2.7 and 4.5.5 from~\cite{BW90}.

The theorem follows immediately from the last claim because a constant 
$\rec{X}{E}$ of \ACP$_\drt^\tau$\REC\ where $E$ is linear recursive 
specification over \ACP$_\drt^\tau$\REC\ is also a constant of 
\BPA$_\drt^\tau$\REC.
\qed
\end{proof}

\section{Analysis of the PAR Protocol Revisited}
\label{sect-analysis-3}

We take up the analysis of the PAR protocol again, but now using the
additional axiom DRB5.

We abstract from the actions in $I$, but not from the timing of actions.
Starting from the specification of 
$\encap{H}(S \parc K \parc L \parc R)$ in
Section~\ref{subsect-specification}, we can now calculate that
$\abstr{I}(\encap{H}(S \parc K \parc L \parc R))$ is the 
$X'''$-component of the solution of the guarded recursive  specification 
that consists of the following equations:
\begin{ldispl}
\begin{aeqns}
X''' & = &
\Altc{d \in D} \cts{r_1(d)} \seqc \delay^{t_S}(Y_d''') \altc
\delay(X''')\;,
\eqnsep
Y_d''' & = &
\delay^{t_K + t_R}(\cts{s_2(d)}) \seqc 
(\delay^{t_R' + t_L}(\cts{\tau} \seqc X''') \altc
 \delay^{t_S' - (t_K + t_R)}(Z'''))
\altc
\delay^{t_S'}(Y_d''')\,,
\eqnsep
Z''' & = &
\delay^{t_K + t_R' + t_L}(\cts{\tau} \seqc X''') \altc
\delay^{t_S'}(Z''')\;.
\end{aeqns}
\end{ldispl}%
Many simplifications have been achieved by using axiom DRB5.
In particular, all internal actions that hinder performance analysis
could be removed.
The ones that could not be removed originate from the communications 
of acknowledgements at port~5 and these may be followed 
by the reception of a datum in the same time slice.

It is also possible to show that the PAR protocol is functionally 
correct by abstracting from the timing of actions next.
That is, we can now calculate, starting from $\drtfp(X''')$, that
$\drtfp(\abstr{I}(\encap{H}(S \parc K \parc L \parc R)))$ is the
solution of the guarded recursive specification of a buffer with
capacity one.  

A more intelligible specification of 
$\abstr{I}(\encap{H}(S \parc K \parc L \parc R))$ can be obtained
if we add to \ACP$_\drt^\tau$\REC, for each $n \in \Nat$, a unary 
operator $\delay^{*n}$ to the signature of \ACP$_\drt^\tau$\REC\ and the
equation $\delay^{*n}(x) = x \altc \delay^n(\delay^{*n}(x))$ and the
conditional equation $y = x \altc \delay^n(y) \Limpl y = \delay^{*n}(x)$
to the axiom system of \ACP$_\drt^\tau$\REC.
By using this extension of \ACP$_\drt^\tau$\REC, the specification of 
$\abstr{I}(\encap{H}(S \parc K \parc L \parc R))$ given above can easily 
be rewritten to the following one:
\begin{ldispl}
\begin{aeqns}
X''' & = &
\Altc{d \in D} 
 r_1(d) \seqc 
  \delay^{* t_S'}(\delay^{t_S + t_K + t_R}(\cts{s_2(d)})) 
\\ & & \phantom{\;\sum\;}
 {} \seqc 
  (\delay^{t_R' + t_L}(\cts{\tau} \seqc X''') \altc
    \delay^{* t_S'}
     (\delay^{t_R' + t_L + t_S' - t_R}(\cts{\tau} \seqc X''')))\;.
\end{aeqns}
\end{ldispl}%
This specification clearly exhibits both the functional behaviour and
the performance properties of the PAR protocol.
It is manifest that the protocol behaves as a buffer with capacity one,
that a datum is delivered $t_S + t_K + t_R + i \cdot t_S'$ time slices
after its consumption (for some $i \geq 0$), and that the next datum 
can be consumed $t_R' + t_L$ or $t_R' + t_L + t_S' - t_R + j \cdot t_S'$ 
time slices after the delivery (for some $j \geq 0$).

\section{Concluding Remarks}
\label{sect-conclusion}

The PAR protocol has been described and analyzed using a version of 
\ACP\ with abstraction for processes with discrete relative timing that
can be considered the core of the versions presented earlier 
in~\cite{BB95b,BBR98a,BM02a}.
The protocol could be described adequately.
In addition, its functional correctness could be analyzed thoroughly.
That is, it could be proved that the protocol behaves correctly if the 
time-out time of the sender is longer than the time that a complete 
protocol cycle takes.
Its performance properties could not be analyzed as thoroughly.
To remedy this, a variant of the standard notion of branching 
bisimilarity for processes with discrete relative timing, called 
dormancy-aware branching bisimilarity, has been proposed.
This variant is coarser than the standard notion, but it still coincides 
with the original notion of branching bisimilarity from~\cite{GW96a} in 
the case of processes without timing. 
A plausible motivation for the variant has been given.
It has also been shown that an additional axiom schema that is valid 
with respect to the variant permits thorough performance analysis of the 
PAR protocol.

In~\cite{Klu93a}, the PAR protocol is described and analyzed using a 
version of \ACP\ with abstraction for processes with continuous absolute 
timing. 
The description originates from an earlier description
in~\cite{BB91a}.
In the case of the analysis in~\cite{Klu93a}, handwavings are needed to 
come as far as in the case of our analysis.
Other protocols that are described and analyzed in versions of \ACP\ 
with abstraction for processes with timing include the ABP (Alternating 
Bit Protocol) and the CABP (Concurrent Alternating Bit Protocol), both 
in~\cite{Hil95a}, and Fischer's mutual exclusion protocol, 
in~\cite{Ver95a}.
In virtually all cases, there is a need for a coarser equivalence.
The equivalences suggested in~\cite{Hil95a,Ver95a} are in the case of
processes without timing also coarser than the original version of 
branching bisimilarity from~\cite{GW96a}.
I claim that, in the case of the above-mentioned protocols, there is 
no need for an equivalence that is coarser than dormancy-aware 
branching bisimilarity.

Dormancy-aware branching bisimilarity is most closely related to the
equivalences that abstract from silent steps introduced for other 
versions of \ACP\ with abstraction for processes with timing.
The equivalences in question are all versions of branching bisimilarity 
for processes with timing.
Dormancy-aware branching bisimilarity is coarser than the version of 
branching bisimilarity for processes with discrete relative timing 
introduced in~\cite{BB95b,BBR98a,BM02a}.
The latter version is in line with the version of branching bisimilarity 
for processes with continuous relative timing introduced 
in~\cite{Klu93a} and both versions are based on the version of branching 
bisimilarity for processes without timing from~\cite{GW96a}. 
In~\cite{Klu93a}, an outline of an unnamed coarser equivalence, denoted
by $\bisim_\mathrm{rb}^*$, is given as well.
That equivalence can be regarded as the first step towards 
dormancy-aware branching bisimilarity: it looks as if its outline 
in~\cite{Klu93a} is the point of departure of the unsatisfactory 
definition of dormancy-aware branching bisimilarity in~\cite{BMR02b}.

It is difficult to compare dormancy-aware branching bisimilarity with 
the equivalences that abstract from silent steps introduced for other 
process algebras for processes with timing, such as the different 
versions of CCS with timing~\cite{Che92a,MT92,Wan90}, 
Timed CSP~\cite{Dav92a}, TIC~\cite{QFA93a} and TPL~\cite{HR95a}.
The reason for it is that they are based on equivalences other than
branching bisimilarity. 
For example, the equivalences for the different versions of CCS with 
timing and TIC are all based on weak bisimilarity, the equivalence for 
Timed CSP is based on failure equivalence, and the different 
equivalences for TPL are based on the different testing equivalences.
For a comparison of those equivalences for processes without timing, 
and many others that abstract from silent steps, the reader is 
referred to~\cite{Gla93a}.

In~\cite{RW07a}, an alternative for dormancy-aware branching 
bisimilarity is proposed.
The underlying idea of that proposal is that abstraction from  
internal actions involves full abstraction from their timing.
This approach leads to rather counterintuitive situations.
For example, with this approach applied to the setting of the current 
paper, we would have that
$\abstr{\set{b}}(\delay(\cts{a}) \leftm \cts{b}) = \dead$ and
$\delay(\cts{a}) \leftm \abstr{\set{b}}(\cts{b}) \neq \dead$ whereas
one intuitively expects to have that
$\abstr{\set{b}}(\delay(\cts{a}) \leftm \cts{b}) =
 \delay(\cts{a}) \leftm \abstr{\set{b}}(\cts{b})$.
Actually, the setting of~\cite{RW07a} is one in which timing is handled 
by time stamping of actions, timing is absolute, and important operators
such as the parallel composition operator are not covered.

As explained in Section~\ref{sect-prelims}, \ACP$_\drt^\tau$ is the core 
of the process algebra \ACP$^\drt_\tau$ presented in~\cite{BM02a}.
Another \ACP-style process algebra closely related to \ACP$_\drt^\tau$ 
is \ACP$_\drt \tau$ from~\cite{BB95b,BBR98a}. 
\ACP$_\drt \tau$ has, in addition to the constants of \ACP$_\drt^\tau$, 
the deadlocked process constant $\didead$ from \ACP$^\drt_\tau$, 
the delayable action constant $a$ for each $a \in \Act$, 
the delayable silent step constant $\tau$, and 
the delayable deadlock constant~$\dead$.
Delayable action constants denote processes that perform an observable 
action in any time slice and after that immediately terminate 
succesfully.
The delayable silent step and delayable deadlock constants can be 
explained analogously.
The axioms defining the delayable action, silent step, and deadlock 
constants make use of an additional unbounded start delay operator 
in~\cite{BB95b} and a time iteration operator in~\cite{BBR98a}.
With the exception of the process denoted by the deadlocked process 
constant, the processes denoted by these constants are definable in
\ACP$_\drt^\tau$\REC.

\bibliographystyle{splncs03}
\bibliography{PA}

\end{document}